\documentclass[12pt]{article}
\usepackage{amsmath,epsf}
\usepackage[T1]{fontenc}
\usepackage[latin1]{inputenc}
\usepackage{epsf}
\usepackage{amsfonts,amssymb}
\usepackage[usenames,dvipsnames]{color}
\newcommand{\de}{\delta}

\newcommand{\ka}{\kappa}
\newcommand{\la}{\lambda}

\newcommand{\vp}{\varphi}
\newcommand{\La}{\Lambda}

\newcommand{\Lao}{\Lambda_0}

\newcommand{\pa}{\partial}

\newcommand{\ti}[1]{\tilde{#1}}

\newcommand{\qed}{\hfill \rule {1ex}{1ex}\\ }

\newcommand{\eq}{\begin{equation}}
\newcommand{\eqe}{\end{equation}}

\newcounter{saveeqn}

\begin{document}

\title{On the local Borel transform of Perturbation Theory}

\author{Christoph Kopper\footnote{\ kopper@cpht.polytechnique.fr} \\
Centre de Physique Th{\'e}orique, CNRS, UMR 7644\\
Ecole Polytechnique\\
F-91128 Palaiseau, France} 

\date{}

\maketitle

\begin{abstract}
We prove existence of the local Borel transform 
for the perturbative series of massive $\vp_4^4$-theory. 
As compared to previous proofs in the literature, the present
bounds are much sharper as regards the dependence on 
external momenta, they are explicit in the number of external 
legs, and they are obtained quite simply through a judiciously
chosen induction hypothesis applied to the Wegner-Wilson-Polchinski
flow equations. We pay attention not to generate an astronomically 
large numerical constant for the inverse radius of convergence 
of the Borel transform.

\end{abstract}

\newpage

\section{Introduction }
Perturbation theory in quantum field theory is suspected to be
divergent. 
The divergent behaviour can be directly related to the presence 
of nontrivial minima of the classical action in the complex coupling
constant plane [Li], and one speaks of instanton singularities
in consequence. 
Starting from the expansion in terms of Feynman diagrams 
the singularity can also be related to the increase of the
number of Feynman diagrams at high orders in perturbation theory.
In theories like $\vp^4$, this number grows as $N!\,$, where N 
is the order of perturbation theory. This indicates divergent
 behaviour. In four dimensions this divergence has never been proven
however. 
The  main obstruction stems from the
 renormalization subtractions which are required to cancel short distance
singularities. They lead to the appearance of
contributions of opposite sign in the Feynman amplitudes.
A lower bound on perturbative contributions would then require
to control the absence of efficient 
sign cancellations, a task which
has turned out to be too difficult up to the present day.  
Thus divergence can only be proven in three or fewer
dimensions where the renormalization problem is marginal or absent
[Sp], [Br],  [MR]. 
In the four-dimensional case the very need for 
renormalization implies the appearance of a  new (hypothetical)
source of divergence      
of the perturbative expansion, named renormalon singularity
after 't Hooft [tH]. This type of singularity is related - in the
language of Feynman graphs - to the presence of graphs which 
require a number of renormalization subtractions proportional to the
order of perturbation theory. In a strictly renormalizable theory it
typically leads to a corresponding power of the  logarithms of the 
momenta flowing through the diagram. 
For example for the diagram of Fig.1 we obtain an integral of the type 
\[
\int d^4p \ \frac{1}{(p^2+m^2)^3}\ \log^N (\frac{p^2+m^2}{m^2}) \ \sim\
N!  \ ,    
\]
where $N\,$ is the number of bubble graph insertions and $p$ the
momentum flowing through the big loop. Such a behaviour is obviously not 
compatible with a convergent perturbation expansion.

\begin{figure}
\begin{center}
\centerline{\mbox{\epsfysize 4cm \epsffile{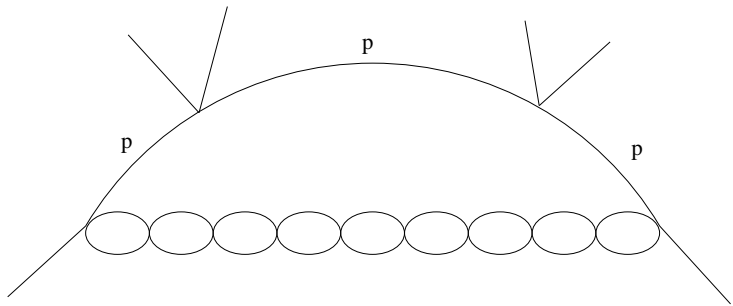}}}
\caption{A renormalon diagram in $\vp_4^4$-theory}
\end{center}
\end{figure}

It was then proven in the seminal work of de Calan and Rivasseau [CR]
that the two sources  of divergent behaviour do not conspire to 
deteriorate the situation even more.     
Even in the presence of both instanton and renormalon type
singularities the Borel transform of the perturbation expansion
has a finite radius of convergence, i.e. perturbative amplitudes 
at order $N\,$ do not grow more rapidly
than $N!\,$. In fact one of the main 
results of  [CR] is that the number of graphs which 
require $k \le N$ renormalization subtractions is bounded by
$(const)^N\ \frac{N!}{k!}\,$ so that the bound they present
on their amplitudes, which is 
of the form $(const')^N\ k!\,$, is sufficient to prove 
local existence of the Borel transform.  

The subject of large orders of perturbation theory was taken up 
by several authors in the sequel. The bounds were  improved 
and generalized in the paper [FMRS]. 
In [CPR] the result was extended to massless $\vp_4^4$-theory.
Local existence of the Borel transform 
for QED  was proven in the book [FHRW]. David, Feldman and Rivasseau 
[DFR] made essential progress in proving that the radius of
convergence of the Borel transformed series for the 
$\vp_4^4$-theory is not smaller
than what is expected from the analysis of typical simple graphs
contributing to the renormalon singularity as the one of Fig.1.
Namely they showed that this radius is bounded below by the inverse 
of the first coefficient of the $\beta$-function, 
as suspected by 't Hooft. In fact this coefficient is
calculated from a subclass  of diagrams of which 
the one shown in Fig.1 is a representative.
They are obtained by iteratively replacing in all possible 
ways elementary vertices by the one-loop bubble graph which apppears 
as a multiple insertion  in Fig.1. 
The proof required a judicious partial resummation
technique applied to the perturbative expansion, of a similar kind as
the one employed previously in [Ri] to prove the existence 
(beyond perturbation theory) of planar "wrong" sign 
$\vp_4^4$-theory. It also made use of the precise upper bounds 
 on the perturbative series in the absence of 
renormalon type diagrams established previously in [MR] and [MNRS].
Finally
Keller [Ke] first proved the local existence of the Borel 
transform in the framweork of the Wegner-Wilson-Polchinski 
flow equations which we also use in this paper.

As compared to the previous papers our motivation and in
 consequence the results are 
different. Our paper is of course closest in spirit to [Ke], 
which is the only one where the dependence on the number of 
external legs is explicitly controlled.
The paper is part of a larger program to get rigorous control 
of the properties of the Schwinger or Green functions of
quantum field theory with the aid of flow equations. A review is in
[M\"u], for  recent novel results see e.g. [KM\"u], [Ko].
Our aim is not only to  control the large order
behaviour of perturbation theory in the sense of the mathematical 
statement on the existence of the local Borel transform.
We would  like to control the whole set of Schwinger functions 
at the same time as regards their large momentum behaviour.
This is in fact necessary  if the bounds on the
Schwinger functions are supposed to serve as an ingredient to
further analysis. If for example they appear as an input in the flow
equations, or simliarly in Schwinger-Dyson type equations,
bad bounds on one side will typically undermine good ones on
the other side~; for example bad high momentum behaviour will lead to
bad high order behaviour when closing loops and integrating over loop 
momenta. In the same way, since an $n$-point function can be otained by merging
two external lines and forming a loop in an $(n+2)$-point function,
bounds which are not sufficiently strong as regards the dependence
on $n$, will not be of much use either.
We need bounds on the high momentum behaviour 
which do not increase faster than logarithmically with momentum
(apart from the two-pont function),
and which are thus optimal for the four-point function,
 in the sense that they are saturated by certain individual   
Feynman amplitudes. Such bounds were proven in [KM], however
without control on the behaviour at large orders of perturbation
theory or at large number of external legs.
In the above cited papers the control on the high momentum behaviour
is far from sufficient, in  [CR] and in [Ke] the radius of convergence   
of the Borel transform shrinks as an inverse power of momentum, in the
other papers the result is not framed in momentum space 
but rather in distributional sense making use of various norms,
and certainly
too far from optimal to be used in the above described context.
We note  that bounds in position space, if optimal
in the above sense, could serve as well as those in momentum
space. We adressed the problem in momentum space here since 
it is of more common use in short distance physics. 
For work with flow equations in position space see [KM\"u]. 

We would also like to stress the fact that we pay much attention 
to the fact not to produce astronomical\footnote{an astronomical
 constant would be one of the form $10^n$ where $n$ is a large
 integer. Our aim is  to show that a small value of
$n\,$ can be accommodated for.} 
constants in the lower bounds   
on the inverse 
radius of convergence of the Borel transformed Schwinger functions.
The paper could have been considerably shortened without that
effort, and the reader will easily find his shortened path through
the paper, if he is not interested in that aspect. 
The constants obtained in the literature are typically astronomically     
large~; in some restricted sense this is even true for the optimal
result [DFR], since the bound obtained is on asymptotically large orders
of perturbation theory, 
allowing smaller orders to be very large. 
In a closed system of equations it is again not possible to relax  
on low orders of perturbation theory without having a drawback on
higher orders. Further considerable effort seems necessary 
if one wants to obtain a close to realistic value for this 
inverse radius. It requires more explicit calculations in lowest
orders which are of course doable since the flow equations provide
an explicit calculational scheme.

Our paper is organized as follows. We first present the flow equation
framework as we will use it in the proof. Then we collect some elementary
auxiliary bounds which are to be used in the proof of the  subsequent
proposition. 
This part could be considerably shortened, were it not for
the above mentioned aim to avoid the appearance of astronomical constants.
Then we present our results and their proof. 
The reader familiar with the domain will realize that the proof is 
comparatively short and (hopefully) transparent. The hardest part of the work 
consisted in finding out the pertinent induction hypothesis.  
   
\section{The flow equation framework}
\vskip.2cm \noindent
Renormalization theory based on the  flow equation (FE) [WH]
of the renormalization group [Wi]
has been exposed quite often in
the literature [Po], [KKS], [M\"u]. So we will introduce it rather  shortly.
The object studied is the regularized generating functional
$L^{\La,\Lao}$ of connected (free propagator) amputated 
Green functions (CAG).
The upper indices  
$\La$ and $\Lao$ enter through the regularized propagator
\[
C^{\La,\Lao}(p)\,=\, {1 \over  p^2+m^2} 
\{ e^{- {p^2+m^2 \over \Lao^2}} -e^{- {p^2+m^2 \over \La^2}} \}
\]
or its  Fourier transform 
\eq
\hat{C}^{\La,\Lao}(x)= 
\int_p C^{\La,\Lao}(p) \,e^{ipx}\ ,\quad  \mbox{ with }\ 
\int_p\,:= \int_{\mathbb{R}^4} 
 {d^4 p \over (2\pi)^4}\ .
\label{di}
\eqe
We assume 
$\,
0 \le \La \le \Lao \le \infty
\,$ 
so that the Wilson flow parameter $\La$ takes the role of an 
infrared (IR) cutoff\footnote{Such a cutoff is of course not necessary
in a massive theory. The IR behaviour is only modified for $\La$ above
$m$.}, whereas  $\Lao$ is the ultraviolet (UV)
re\-gularization. The full propagator is recovered for 
$\La=0$ and $\Lao \to \infty\,$.
For the "fields" and their Fourier transforms we write 
$\,
\hat{\vp}(x) = \int_p  \vp(p) \ e^{ipx}\,$, $\,
{\de \over \de\hat{\vp}(x)} =
(2\pi)^4 
\int_p {\de \over \de \vp(p)}\, e^{-ipx}\,$.
For our purposes the fields $\hat{\vp}(x)\,$ may be assumed to live in
the Schwartz space ${\cal S}(\mathbb{R}^4)$.
For finite $\Lao$ and in finite volume the theory can be given
rigorous meaning starting from the functional integral
\eq
e^{-\frac{1}{\hbar}(L^{\La,\Lao}(\hat{\vp})+ I^{\La,\Lao})}
\,=\, 
\int \, d\mu_{\La,\Lao}(\hat{\phi}) \; 
e^{- \frac{1}{\hbar}L_0(\hat{\phi}\,+\,\hat{\vp})} \ .
\label{funcin}
\eqe
On the r.h.s. of 
(\ref{funcin}) $\,d\mu_{\La,\Lao}(\hat\phi) $ denotes the (translation
invariant) Gaussian measure with covariance $\hbar \hat{C}^{\La,\Lao}(x)$.
The functional $L_0(\hat\vp)$ is the bare action including
counterterms, viewed as a formal power
series in $\hbar\,$. Its general form for 
symmetric $\vp_4^4$
theory is 
\[
 L^{\Lao,\Lao}(\hat{\vp}) 
= {g \over 4!}  \int \! \! d^4 x \, \hat{\vp}^4(x)  
   \; + 
\]  
\eq
   + \int \! \!d^4 x \,\{{1 \over 2}\, a(\Lao)\ \hat{\vp}^2(x) +
    {1 \over 2}\, b(\Lao)\, \sum_{\mu=0}^3 (\pa_{\mu}\hat{\vp})^2(x) +
    {1 \over 4!}\,c(\Lao)\ \hat{\vp}^4(x)\} \ ,
\label{nawi}
\eqe  
the parameters $a(\Lao),\ b(\Lao),\ c(\Lao)\,$
fulfill
\eq
a(\Lao),\ c(\Lao) =O(\hbar)\,,\quad b(\Lao) =O(\hbar^2)\ .
\label{con}
\eqe
They are directly related to the standard mass, coupling constant and
wave function counterterms. 
On the l.h.s. of 
(\ref{funcin}) there appears the normalization factor 
$\,e^{-I^{\La,\Lao}} $  which is
due to vacuum contributions. The exponent $I^{\La,\Lao}\,$
diverges in infinite volume so that we can take the infinite volume
limit only when it does not appear any more.
We do not make the finite
volume explicit here since it plays no role in the 
sequel. For a more thorough
discussion see [M\"u], [KMR].

The FE is obtained from (\ref{funcin}) on differentiating 
w.r.t. $\La\,$. It is a differential equation for the functional
$L^{\La,\Lao}\,$~: 
\eq
\partial_{\La}(L^{\La,\Lao} + I^{\La,\Lao} )\,=\, 
\label{fe}
\eqe
\[
\,=\,\frac{\hbar}{2}\,
\langle\frac{\delta}{\delta \hat\vp},(\partial_{\La}  \hat C^{\La,\Lao})
\frac{\delta}{\delta \hat\vp}\rangle L^{\La,\Lao}
\,-\,
\frac{1}{2}\, \langle \frac{\delta}{\delta
  \hat\vp} L^{\La,\Lao},(\partial_{\La}
 \hat C^{\La,\Lao})
\frac{\delta}{\delta \hat\vp} L^{\La,\Lao}\rangle \ .
\]
By $\langle\ ,\  \rangle$ we denote the standard scalar product in 
$L^2(\mathbb{R}^4, d^4 x)\,$. Changing to momentum space and
expanding in a formal powers series w.r.t. $\hbar\,$ we write
\[
L^{\La,\Lao}(\vp)\,=\,\sum_{l=0}^{\infty} \hbar^l\,L^{\La,\Lao}_{l}(\vp)\,.
\]
From $L^{\La,\Lao}_{l}(\vp)$ we then define the CAG of order $l$
in momentum space  through 
\eq
\de^{(4)} (p_1+\ldots+p_{n})\, {\cal L}^{\La,\Lao}_{n,l}
(p_1,\ldots,p_{n-1})\,=\,
\frac{1}{n!}(2 \pi)^{4(n-1)} \de_{\vp(p_1)} \ldots \de_{\vp(p_n)}
L^{\La,\Lao}_l|_{\vp \equiv 0}\ \ ,
\label{cag}
\eqe
where we have written 
$\delta_{\vp(p)}=\delta/\delta\vp(p)$.
The CAG are symmetric  in their momentum arguments by definition.
Note that by our definitions the free 
two-point function is not contained  in $L^{\La,\Lao}_{l}(\vp)\,$,
since it is attributed to the Gaussian measure in (\ref{funcin}).
This is
important for the set-up of the inductive scheme, from which we will 
prove our bounds below. 
We thus define
\[
{\cal L}^{\La,\Lao}_{2n,l}\equiv 0 \ \mbox{ for } l < 0\,, \ n \ge 1\ ,\quad
\mbox{and}\quad {\cal L}^{\La,\Lao}_{2,0}\equiv 0 \  .
\] 
The FE (\ref{fe}) rewritten in terms of the CAG (\ref{cag})
takes the following form
\eq
\pa_{\La} \pa^{w} \, {\cal L}^{\La,\Lao}_{2n,l}(p_1,\ldots p_{n-1}) =
({2n+2 \atop 2}) \int_k (\pa_{\La}{C}^{\La,\Lao}(k))\,\pa ^w 
{\cal L}^{\La,\Lao}_{2n+2,l-1}(k,-k, p_1,\ldots p_{2n-1})
\label{feq}
\eqe
\[
-\!\!\!\!\!\!\!\!\!\!\!\!
\sum_{\begin{array}{c}_{l_1+l_2=l,}\atop
_{w_1+w_2+w_3=w}\\[-.15cm]
_{n_1+n_2=n +1}\end{array} }\!\!\!\!\!\!\!\! \!\!\!\!2\, n_1\,n_2 \ 
c_{\{w_j\}}\Biggl[ \pa^{w_1} {\cal
  L}^{\La,\Lao}_{2n_1,l_1}(p_1,\ldots,p_{2n_1-1})\,
\,(\pa^{w_3}\pa_{\La} {C}^{\La,\Lao}(q))\,\,
\pa^{w_2}{\cal L}^{\La,\Lao}_{2n_2,l_2}(p_{2n_1},\ldots,p_{2n-1})
\Biggr]_{sy}
\]
\[ \mbox{with }\quad
q= -p_1 -\ldots -p_{2n_1-1}\,= \,- \,p_{2n_1}
\,= \,p_{2n_1+1} +\ldots +p_{2n}\ .
\]
Here we have written (\ref{feq})  directly in a form
where also momentum derivatives of the  CAG (\ref{cag})
are performed. In this paper we will restrict for simplicity 
to up to 3 derivatives all taken w.r.t.
{\it one} momentum $p_i\,$, since our aim is in the first place 
to bound the Schwinger functions themselves, and not their derivatives
\footnote{In distributing the derivatives over the three
  factors in the second term on the r.h.s. with the Leibniz rule,
we have tacitly assumed
  that the momentum $p_i$ appears among those from 
${\cal  L}^{\La,\Lao}_{2n_1,l_1}\,$. If this is not the case 
one has to parametrize  ${\cal L}^{\La,\Lao}_{2n_1,l_1}\,$
in terms of (say) $(p_2,\ldots p_{2n_1})\,$ with 
$\,p_{2n_1}\,= \,-p_{2n_1+1}-\ldots -p_{2n}\,$, to introduce the 
 $p_i$-dependence in   $ {\cal L}^{\La,\Lao}_{2n_1,l_1}\,$.
For an extensive  systematic treatment including the general situation 
where derivatives w.r.t. several external  momenta are present, see [GK]. 
This situation, also considered in [KM], could be 
analysed here too at the prize of basically notational complication.}. 
We use the shorthand\footnote{slightly abusive, since the index $i$ is
suppressed in $w$} notations 
\[
\pa^w:= \prod_{\mu=0}^{3}
({\pa \over \pa p_{i,\mu}})^{w^{\mu}}\ \mbox{ with }\
w=  (w^{0},\ldots,w^{3})\ , \quad
|w|=\sum_{\mu} w^{\mu}
\]
and 
\[
w!\,=\, w^{0}!\ldots w^{3}!\ , \quad
c_{\{w_j\}}\,=\, \frac{w!}{w_1!\,w_2!\,w_3!} \ .
\] 
The symbol $sy$ 
means taking the mean value over those permutations $\pi\,$
of $(1,\ldots, 2n)\,$, for which 
$\pi(1) < \pi(2) <\ldots < \pi(2n_1-1)\,$ and 
$\,\pi(2n_1) < \pi(2n_1+1) < \ldots < \pi(2n)\,$.\\
For the derivatives of the propagator we find the following relations
\eq
\pa_{\La} C^{\La,\Lao}(p) = -\frac{2}{\La ^3} \
e^{-\frac{p^2+m^2}{\La ^2}}\ ,\quad
\pa_{p_{\mu}}e^{-\frac{p^2+m^2}{\La ^2}}
= -\,\frac{2\,p_{\mu}}{\La ^2} \
e^{-\frac{p^2+m^2}{\La ^2}}\ ,
\label{33}
\eqe
\eq
\pa_{p_{\mu}}\pa_{p_{\nu}} e^{-\frac{p^2+m^2}{\La ^2}} = 
\Bigl[\frac{4}{\La ^4}\ 
p_{\mu}\,p_{\nu}\,-\, \frac{2}{\La ^2}\ \de_{\mu \nu}\Bigr]
e^{-\frac{p^2+m^2}{\La ^2}}\ ,
\label{34}
\eqe
\eq
\pa_{p_{\mu}}\pa_{p_{\nu}}\pa_{p_{\rho}}e^{-\frac{p^2+m^2}{\La ^2}} = 
\Bigl[-\,\frac{8}{\La ^6}\ p_{\mu}\,p_{\nu}\,p_{\rho}\,+\,
\frac{4}{\La ^4} \, \bigl(\de_{\mu \nu} p_{\rho}\,+\,
\de_{\mu \rho} p_{\nu}\,+\,
\de_{\nu \rho} p_{\mu}\bigl)\Bigr]\
e^{-\frac{p^2+m^2}{\La ^2}}\ .
\label{35}
\eqe

\section{A collection of elementary bounds }
The subsequent lemmata state a number bounds which we will make
recurrent use of in the proof of our main result.\\[.1cm] 

\noindent
{\it Lemma 1}~: For $l \in \mathbb{N}_0\,$\\ 
a)
\eq
\sum_{0\le l_1,l_2,\atop l_1+l_2 =l} \frac{1}{(l_1+1)^2\, (l_2+1)^2} 
\ \le \
\frac{5}{(l+1)^2}\ , \qquad 
\sum_{1\le l_1,l_2,\atop l_1+l_2 =l} \frac{1}{(l_1+1)^2\, (l_2+1)^2} 
\ \le \
\frac{3}{(l+1)^2}\ ,\label{ineq12}
\eqe
b)
\eq
\sum_{1\le n_1,n_2,\atop n_1+n_2 =n+1}  \frac{1}{n_1^3\, n_2^3} 
\ \le \
\frac{4}{n^3}\ ,\quad
\sum_{2\le n_1,n_2,\atop n_1+n_2 =n+1}  \frac{1}{n_1^3\, n_2^3} 
\ \le \
\frac{2}{n^3} \ . 
\label{ineq13}
\eqe
\\[.1cm]
{\it Proof}~:
a) The inequality can be verified explicitly for $l\le 5$.
Assuming $l >5$ we have 
\eq
\sum_{0\le l_1,l_2,\atop l_1+l_2 =l} \frac{1}{(l_1+1)^2\, (l_2+1)^2} 
\,=\, \frac{2}{(l+1)^2}\ +\  
\sum_{k=1}^{l-1} \frac{1}{(k+1)^2\, (l-k+1)^2}
\label{16}
\eqe
\[
\le  \frac{2}{(l+1)^2}\, +\,
\int_0^{l}\frac{dx}{(x+1)^2(l-x+1)^2}\,=\, \frac{2}{(l+1)^2}\, +\,
\int_1^{l+1}\!\!\!\!\!
 dx \Bigl(\frac{a+bx}{x^2}+ \frac{c-bx}{(l+2-x)^2}\Bigr)\ ,
\]
where 
\[
a\,=\,\frac{1}{(l+2)^2}\ ,\quad 
b\,=\,\frac{2}{(l+2)^3}\ ,\quad
c\,=\,\frac{3}{(l+2)^2}\ . 
\]
The integral equals then
\eq
\frac{1}{(l+2)^2}\, \Bigl(2[1-\frac{1}{l+1}]\,+\,  
\frac{4}{l+2}\, \log(l+1)\Bigr)\, \le \frac{3}{(l+1)^2}\quad \mbox{for }\
l >5\  ,
\label{int12}
\eqe 
and the bound is thus also verified for $l >5\,$.
The second statement in (\ref{ineq12}) is a direct consequence of the first
 since a term $\,\frac{2}{(l+1)^2} \,$ is subtracted on the l.h.s. 
\\ 
b) We may again assume $n> 5\,$ on verifying the lowest values 
explicitly.  
The statement then follows from the proof of a) through
\[
\sum_{1\le n_1,n_2,\atop n_1+n_2 =n+1}  \frac{1}{n_1^3\, n_2^3} 
\ = 
\sum_{0\le n_1,n_2,\atop n_1+n_2 =n-1}  \frac{1}{(n_1+1)^3\, (n_2+1)^3} 
\]
\[
\le \
\frac{2}{n^3} \,+\, \sup_{1\le n_1 \le n-1} \frac{1}{(n_1+1)\, (n-n_1)} 
\sum_{1\le n_1,n_2,\atop n_1+n_2 =n-1}  \frac{1}{(n_1+1)^2\, (n_2+1)^2} 
\]
\[
\le \
\frac{2}{n^3} \,+\, \frac{1}{2(n-1)} 
\sum_{1\le n_1\le n-2}  \frac{1}{(n_1+1)^2\, (n-n_1)^2} 
\ \le\
\frac{2}{n^3} \,+\,  \frac{1}{2(n-1)}\  \frac{3}{n^2}
\ \le\ \frac{4}{n^3} \ ,
\]
where we used the bound (\ref{int12}) on (\ref{16})
in the last but second inequality.
The second inequality in b) then follows directly from the 
previous calculation.
\qed

\noindent
{\it Lemma 2}~:\\
a)  For integers 
$n \ge 3\,$, $\,n_1\,,\, n_2 \ge 1\,$,  
$\ l\,,\, l_1\,,\, \la_1\,,\ l_2\,,\, \la_2 \ge 0\,$  
\[
\sum_{\begin{array}{c}_{l_1+l_2=l},\atop
_{n_1+n_2=n+1},\\_{\la_1 \le l_1, \,\la_2 \le l_2},
\atop _{\la_1+\la_2=\la}
 \end{array} } 
\frac{1}{(l_1+1)^2\, (l_2+1)^2\,n_1^2\,n_2^2}\ 
\frac{n!}{ n_1!\,n_2!}\ 
\frac{\la!}{\la_1!\,\la_2!} \
 \frac{(n_1+l_1-1)!\ (n_2+l_2-1)!}{(n+l-1)!}
\]
\eq
\le\
K_0\ \frac{1}{(l+1)^2}\ \frac{1}{n^2}\ ,\quad 
\mbox{ where we may choose }\quad K_0=20\ .
\label{le5-1}
\eqe
For $\,n_1, n_2\, \ge 2\,$
\[
\sum_{\begin{array}{c}_{l_1+l_2=l},\atop
_{n_1+n_2=n+1},\\_{\la_1 \le l_1, \,\la_2 \le l_2},
\atop _{\la_1+\la_2=\la}
 \end{array} } 
\frac{1}{(l_1+1)^2\, (l_2+1)^2\,n_1^2\,n_2^2}\ 
\frac{n!}{ n_1!\,n_2!}\ 
\frac{\la!}{\la_1!\,\la_2!} \
 \frac{(n_1+l_1-1)!\ (n_2+l_2-1)!}{(n+l-1)!}
\]
\eq
\le\
\frac12\ K_0\ \frac{1}{(l+1)^2}\ \frac{1}{n^2}\ . 
\label{le5-11}
\eqe
\\
b)  For $\,n \ge 3\,$,  $\,n_1=2\,,\ n_2=n-1\,$
\[
\sum_{\begin{array}{c}_{l_1+l_2=l},\atop _{\la_1 \le l_1, \,\la_2 \le
      l_2}, 
\\[-.15cm]_{\la_1+\la_2=\la}
 \end{array} } 
\frac{1}{(l_1+1)^2\, (l_2+1)^2\,n_1^2\,n_2^2}\ 
\frac{n!}{ n_1!\,n_2!}\ 
\frac{\la_1 !}{\la_1!\,\la_2!} \
 \frac{(n_1+l_1-1)!\ (n_2+l_2-1)!}{(n+l-1)!}
\]
\eq
\le\
K_0'\ \frac{1}{(l+1)^2}\ \frac{1}{n^2} \  ,\quad 
\mbox{ where we may choose }\quad  K'_0=  (\frac34)^3\cdot 5\ \le\ 2.2
\ .
\label{le5-2}
\eqe
c)  For $\,n \ge 2\,$, $\,n_1=1\,,\ n_2=n\,$
\[
\sum_{\begin{array}{c}_{l_1+l_2=l},\atop_{\la_1 \le l_1, \,\la_2 \le l_2},
\\[-.15cm]_{\la_1+\la_2=\la}
 \end{array} } 
\frac{1}{(l_1+1)^2\, (l_2+1)^2\,n_1^2\,n_2^2}\ 
\frac{n!}{ n_1!\,n_2!}\ 
\frac{\la!}{\la_1!\,\la_2!} \
 \frac{(n_1+l_1-1)!\ (n_2+l_2-1)!}{(n+l-1)!}
\]
\eq
\le\
 K_0''\ \frac{1}{(l+1)^2}\ \frac{1}{n^2} \  ,\quad 
\mbox{ where we may choose }\quad  K_0''=  5 \ .
\label{le5-3}
\eqe
\\[.1cm]
{\it Proof}~: a) We have 
\[
\frac{ n!}{ n_1!\,n_2!}\,
\frac{\la!}{\la_1!\,\la_2!}\, 
 \frac{(n_1+l_1-1)!\ (n_2+l_2-1)!}{(n+l-1)!}
\, =\, \frac{n}{n_1\,n_2}\, 
\bigl({n\!-\!1\atop n_1\!-\!1}\bigr)\, \bigl({\la\atop \la_1}\bigr)\,
\Bigr[\bigl({n+l\!-\!1 \atop n_1+l_1\!-\!1}\bigr)\Bigr]^{-1}\ .
\]
We note that 
\eq
\bigl({n\!-\!1\atop n_1\!-\!1}\bigr)\ \bigl({l\atop l_1}\bigr)\
\le \ 
\bigl({n+l\!-\!1\atop n_1+l_1\!-\!1}\bigr)\ .
\label{b}
\eqe
This follows directly from the standard identity
\[
\sum_{k=0}^{p}  
\bigl({n\!-\!1\atop p-k}\bigr)\ \bigl({l \atop k}\bigr)\
=\
\bigl({n +l\!-\!1\atop p}\bigr)\ ,
\]
assuming without limitation that $n-1 \ge l\,$ and
setting $p = \inf\{n_1+l_1-1, n+l-(n_1+l_1)\} 
\le \frac{n+l-1}{2}\le n-1\,$.\\
Secondly we show that for $\,l=l_1+l_2$
\eq
\sum_{\begin{array}{c}_{\la_1 \le l_1, \,\la_2 \le l_2},
\\[-.15cm]_{\la_1+\la_2=\la}
 \end{array} }\frac{\la!}{\la_1!\,\la_2!}\
\ \le \ \bigl({l\atop l_1}\bigr)\ .
\label{c}
\eqe
For the inductive proof we assume $ l \ge 1\,$ and 
 without loss  $\, l_2\le l_1\,$ . To realize
by induction on $\, 0\le k \le l_2\,$ that
\[
A_k~:=  \Bigl[\bigl({l\atop l_1}\bigr)\Bigr]^{-1}\
\sum_{\begin{array}{c}_{\la_1 \le l_1, \,\la_2 \le l_2},
\\[-.15cm]_{\la_1+\la_2=l-k} \end{array} }
\frac{(l-k)!}{\la_1!\,\la_2!}\ \le 1\ ,
\]
we start from $A_0=1\,$. 
Then assuming that  we have $\,A_{k-1} \le 1\,$  for $k \ge 1\,$ 
we find
\[
A_{k}\, =\,
\frac{l_1-(k-1)}{l-(k-1)}\, A_{k-1}\ +\
\Bigl[\bigl({l\atop l_1}\bigr)\Bigr]^{-1}\
\bigl({l-k\atop l_1}\bigr)
\, \le \,
1- \frac{l_2}{l-(k-1)}\ +\  \frac{l_2}{l}\, 
\frac{(l_2-1)\ldots (l_2-(k-1))}{(l-1)\ldots (l-(k-1))}\ .
\]
This equals $1$ for $k=1$ and can be bounded for $k\ge 2$ through
\[
1- \frac{l_2}{l-(k-1)}\bigl(1-  
\frac{(l_2-1)(l_2-2)\ldots (l_2-(k-1))}
{\quad \ l\quad \ \, (l-1)\,\ldots\ (l-(k-2))}\bigr)\ \le\ 1\ .
\]
For $\,l_2\, < \, k \, \le\, l\,$ it is immediate to see
that  $\,A_{k} \le A_{k-1}\,$ since the sum for  $\,A_{k} \,$ 
does not contain more nonvanishing terms than the one for  $\,A_{k-1}
\,$, and a nonvanishing term in  $\,A_{k} \,$ can be
bounded by a corresponding one in   $\,A_{k-1}\,$~:
\[
\frac{(l-k)!}{\la_1!\,\la_2!}\ \le \
\frac{(l-(k-1))!}{(\la_1+1)!\,\la_2!}\ .
\]
Now it follows from (\ref{b}), (\ref{c}) that 
\eq
\sum_{\begin{array}{c}_{\la_1 \le l_1, \,\la_2 \le l_2},
\\[-.15cm]_{\la_1+\la_2=\la}
 \end{array} }\frac{n}{n_1\,n_2}\
\frac{(n_1+l_1-1)!}{(n_1-1)!\,\la_1!}\
\frac{(n_2+l_2-1)!}{  (n_2-1)!\,\la_2!}\
\frac{(n-1)!\,\la!}{(n-1+l)!} \ \le \ \frac{n}{n_1\,n_2}\ .
\label{ni}
\eqe
Using Lemma 1 we then get
\eq
\sum_{\begin{array}{c}_{l_1+l_2=l},\\[-.15cm]
_{n_1+n_2=n+1}
 \end{array} } \!\!\!\! \frac{n\,}{n_1\,n_2}\ 
\frac{1}{(l_1+1)^2\, (l_2+1)^2\,n_1^2\,n_2^2}\ 
\le\
\frac{20}{(l+1)^2\, n^2} \ .
\label{endle2}
\eqe
The statements (\ref{le5-11}) and parts b) (\ref{le5-2}) and  c) 
(\ref{le5-3}) follow from Lemma 1 and (\ref{ni}).
\qed

\noindent
{\it Lemma 3}~:
For $v \le 3\,$ and  $a_i\,,\, x\in \mathbb{R}^4\,$
the following inequality holds
\begin{equation}
  e^{-\frac{x^2}{2}}\ \prod_{i=1}^{v} 
\frac{1}{\sup(1,|x+a_i|)}\,\leq\,c(v)\,
\prod_{i=1}^{v}
\frac{1}{\sup(1,|a_i|)}\ ,
\label{ineq1}
\end{equation}
where we may choose
\eq
c(0) \,=\,1\,, \quad  
c(1) \,=\,1.4\,,\quad  c(2) \,=\,2.5\,,\quad  c(3) \,=\,5.25\ .
\label{coef}
\eqe 
\\[.1cm]
{\it Proof}~: The inequality is trivial
if one allows for large constants.
Suppose $\,v=3\,$. We may suppose without limitation that $|a_3|\ge
|a_2|\ge |a_1| \ge 1\,$ (if $a_i \le 1$ we may pass to the case  
$\,v-1\,$), 
and that $|x| \le \sup |a_i|\,$ since the expression on the
  l.h.s. of  (\ref{ineq1}) is maximized
if all $\,a_i \in \mathbb{R}^4\,$ are parallel and  anti-parallel to
$\,x\,$. 
In this case, assumming that  $\,|a_3|\,|a_2|\,|a_1|\,\ge (1+|x|)^3$,
the inequality at fixed product $\,|a_3|\,|a_2|\,|a_1|\,$
and at fixed $|x|$, 
becomes most stringent if $|a_1|,\,|a_2|=1+|x|\,$.
It then takes the form
\eq
  e^{-\frac{x^2}{2}}\ (1+|x|)^2\,\leq\,c(3)\,
\frac{|a_3|-|x|}{|a_3|}\quad  \mbox{ with }\quad |a_3| >1+|x|
\ .
\label{leid}
\eqe 
If  $\,|a_3|\,|a_2|\,|a_1|\,<\, (1+|x|)^3\,$, 
the bound is satisfied if we demand
\[
  e^{-\frac{x^2}{2}}\,\leq\,c(3)\,
\frac{1}{(1+|x|)^3}
\ .
\] 
This relation is also sufficient for (\ref{leid}) to hold. 
The expression $\,  e^{-\frac{x^2}{2}}\ (1+|x|)^3\,$ 
is maximal for $|x|= \frac{\sqrt{13}-1}{2}\,$
and bounded by $5.25\,$.
The cases $\,v=2\,$ and $\, v=1\,$ are treated analogously.\qed

\noindent
{\it Lemma 4}~: For $r \in \mathbb{N}\,$ and $a \ge 0\,$
\eq
\int_x e^{-\frac{|x|^2}{2}}\ \log^{r}(|x|+a)\ \le \
\frac14\, \log_{+}^r a \,+\, \frac13 \,(r!)^{1/2}\ , 
\label{ineq6}
\eqe
where $\log_{+} x~:= \log(\sup(1,x))\,$.\\[.1cm]
{\it Proof}~: Again the only nontrivial point is to avoid  bad numerical
constants in the bound.  
Remembering the definition (\ref{di}), first note that
for $\,r \le 6 \,,\ a \le 3$
\[
\int_{x}   e^{-\frac{|x|^2}{2}}\ \log^{r}(|x|+a) \ \le\ 
\int_{x}   e^{-\frac{|x|^2}{2}}\ \log^{r}5 \,+\,
\int_{|x| \ge 2 }   e^{-\frac{|x|^2}{2}}\ \log^{r}(|x|+a) 
\]
\eq
\le\ \frac{(1.61)^r}{4\pi ^2}\ +\ \frac{1}{8\pi ^2}
\sum_{n \ge 2} e ^{-n^2/2}\ n^3\ \log^r(3+n)\ \le \
\frac{1}{3}\ (r!)^{1/2} \quad \surd
\label{ineq4}
\eqe
on bounding the sum numerically~; 
we also used the fact that the derivative of
the integrand w.r.t. $|x|$ is negative for $|x| \ge 2\,$.
Secondly, for $\,r \le 6 \,,\ a > 3$
\eq
(\log a)^{-r} \int_{x}  e^{-\frac{|x|^2}{2}}\ \log^{r}(|x|+a)\ =\
\int_{x} e^{-\frac{|x|^2}{2}}\ 
\bigl[1\,+\, \frac{\log (1+\frac{|x|}{a})}{\log a}\bigr]^r 
\label{102}
\eqe
\[
\le \ \int_{x}  e^{-\frac{|x|^2}{2}}\ (1+\frac{\log 2 }{\log a})^r\ +\
\frac{1}{8 \pi ^2}\sum_{n \ge 3}  e ^{-n^2/2}\ n^3
\bigl[1\,+\, \frac{\log (1+\frac{n}{3})}{\log 3}\bigr]^6
\ \le  \ \frac{1}{4\pi ^2}\ (1+\frac{\log 2 }{\log a})^r
\ +\  \frac{6.5}{8 \pi ^2}
\]
\[
\le \ \frac14\ +\ \frac13\, \frac{(r!)^{1/2}}{\log^r a}\quad \surd
\]
on bounding the sum numerically and on noting that 
the last inequality is valid taking $a=3$ on the l.h.s.
and $a=5$ on the r.h.s., and also for $a = 5$ on the l.h.s.
and $a= e ^2$ on the r.h.s. For $\log a \ge 2$
the last bound can be replaced by $\frac14\,$ independently of $ r \le 6\,$. 
Thirdly,   for   $\,r > 6 \,,\ a \le r$
\[
\int_x e^{-\frac{|x|^2}{2}}\ \log^{r}(|x|+a)\ \le \
\int_x e^{-\frac{|x|^2}{2}}\ \log^{r}(|x|+r) 
\ \le \
\log^{r}r\int_x  e^{-\frac{|x|^2}{2}}\ 
\bigl[1\,+\, \frac{|x|}{r\, \log r}\bigr]^r 
\]
\[
 \le \
\log^{r}r\int_x  e^{-\frac{|x|^2}{2}+ \frac{|x|}{\log r}}\ 
\le \ 
\log^{r}\!r 
\ \frac{e ^{\frac{1}{2\log^2 6}}}{4\pi ^2} 
\int_{-\frac{1}{\log 6}}^{\infty}  
e^{-z}\ z\ dz \ \le \ \frac{1}{10} \ \log^{r}r 
\]
on majorizing for $r=6\,$ 
and completing the square in the last but second integral.
Then 
\[
\frac{1}{10} \ \log^{r}r \le \frac{1}{3}\ (r!)^{1/2}\ ,
\]
noting that $\log^{r}r/ (r!)^{1/2} \le 2.75\,$, the maximal value being
attained for $r=15$.\\   
In the fourth place we have for $\,a > r > 6$ quite similarly
\[
\frac{1}{\log^{r}a }\int_x  e^{-\frac{|x|^2}{2}}\ 
\log^r(|x|+a) \, = 
\int_x  e^{-\frac{|x|^2}{2}}\ 
\bigl[1\,+\, \frac{\log (1+\frac{|x|}{a})}{\log a}\bigr]^r
\, \le \int_x  e^{-\frac{|x|^2}{2}}\ 
\bigl[1\,+\, \frac{|x|}{r\, \log r}\bigr]^r 
\, \le \,
\frac{1}{10}  
\ .
\]
\qed

\noindent
{\it Lemma 5}~: 
For $s \in \mathbb{N}\,,\ a >0\,,\ M >\ka \ge m >0\,$
\eq
\sum_{\la=0}^{\la =l}\frac{1}{2^{\la}\,\la!}\ 
\int_\ka^{M} d\ka'\ \ka'^{-s-1} \ 
\log^{\la}(\sup({a \over \ka'},{\ka'\over m}))
\ \le\
3\ \frac{\ka^{-s}}{s}\, 
\sum_{\la=0}^{\la =l} \frac{1}{2^{\la}\,\la!}\,
\log^{\la}\sup({a \over \ka},{\ka \over m})\ .
\label{weh1}
\eqe
{\it Proof}~:
We have
\[
\int_\ka^{M} d\ka'\ \ka'^{-s-1} \ 
\log^{\la}(\sup({a \over \ka'},{\ka'\over m}))
\ \le
\frac{\ka^{-s}}{s}\, 
\log_+^{\la}({a \over \ka})
\, +\,
\int_{\sup(\ka, \sqrt{m a }\,)}
^{\sup(\sqrt{m a} ,\,M)} 
d\ka'\ \ka'^{-s-1} 
\,\log^{\la}(\frac{\ka'}{m})
\ ,
\]
and the last integral can be bounded by
\eq
\int_{\ka}^{M} 
d\ka'\ \ka'^{-s-1} 
\ \log^{\la}(\frac{\ka'}{m})
\ \le \ \frac{\ka^{-s}}{s}\
\la!\,\sum_{\nu  =0}^{\la}\frac{\log^{\nu}(\frac{\ka}{m})}{\nu!}\
\frac{1}{{s}^{\la-\nu}}\ .
\label{fac}
\eqe
We then find
\[
\sum_{\la=0}^{\la =l}\Bigl\{ \frac{1}{2^{\la}\,\la!}\ 
\log_+^{\la}({a \over \ka})
\ +\
 \frac{1}{2^{\la}}\ 
\sum_{\nu  =0}^{\la}\frac{\log^{\nu}(\frac{\ka}{m})}{\nu!}\
\frac{1}{{s}^{\la-\nu}}\Bigr\}
\]
\[
\le \ \sum_{\la=0}^{\la =l} 
\frac{\log^{\la}({a\over \ka})}{2^{\la}\,\la!}
\,+\,2\, \sum_{\la=0}^{\la =l} 
\frac{\log^{\la}({\ka \over m})}{2^{\la}\,\la!}
\, \le\,
3\,\sum_{\la=0}^{\la =l} \frac{1}{2^{\la}\,\la!}\,
\log^{\la}\sup({a \over \ka},{\ka \over m})\ .
\]
\qed

\noindent
{\it Lemma 6}~:\\[.1cm] 
Here and in the following we set $\ka=\,\La+m\,$.\\
a)
\eq 
\frac{2}{\La ^3} \
e^{-\frac{m^2}{\La ^2}}\ \le \
\frac{ K_2}{\ka ^3}\ ,\quad\ \mbox{ where }\
K_2 = 6.2\  ,
\label{la3}
\eqe
\eq
p^{2}\ e^{-\frac{p^2}{2\La ^2}}\ \le \
\ka ^2\ \frac{2}{e}\ ,
\qquad  
|p|\ e^{-\frac{p^2}{2\La ^2}}\ \le \
\ka \ \frac{1}{\sqrt{e}}\ .
\label{p}
\eqe
\\[.1cm]
b)
${\ \ \; }$ For $\,|w| \le 3\,$: 
\eq
|\pa ^{w}\,\frac{2}{\La ^3}\,e^{-\frac{p^2}{\La ^2}}
\,e^{-\frac{m^2}{\La ^2}}|\ \le \
K^{(|w|)}\ \ka ^{-3}\
[\sup(\ka, |p|)]^{-|w|} \ .
\label{kw1}
\eqe
\centerline{with $\ \ \ K^{(0)}= K_2\,, \,
\ K^{(1)}=  \frac{2 K_2}{e}\,=\,4.6\,,\ \,
K^{(2)}= 77.5\,,\ \, K^{(3)}= 37\,$.}
\\[.1cm]
\eq
|\pa ^{w}\,\frac{2}{\La ^3}\,e^{-\frac{p^2}{2\La ^2}}
\,e^{-\frac{m^2}{\La ^2}}|\ \le \
K'^{(|w|)}\ \ka ^{-3}\
[\sup(\ka, |p|)]^{-|w|} \ .
\label{kw}
\eqe
\centerline{with $\ \ \ K'^{(0)}= K_2\,, \,
\ K'^{(1)}=  \frac{4 K_2}{e}\,=\,9.2\,,\ \,
K'^{(2)}= 135\,,\ \, K'^{(3)}= 407\,$.}
\\[.1cm]
c) For $\,0 \le \tau \le 1\,$ and $\,p_4(\tau)=-\tau p_1-p_2- p_3\,$: 
\eq 
\frac{ |p_1|\ e^{-\frac{p_1^2}{2\La ^2}} }
{\sup(\ka,\eta_{1,4}^{(4)}(\tau p_1,p_2,p_3,p_4(\tau)))}\
\le \ 
e^{-1/2} \ ,\quad
\frac{ |p_1^2|\ e^{-\frac{p_1^2}{2\La ^2}} }
{\sup(\ka,\eta_{1,4}^{(4)}(\tau p_1,p_2,p_3,p_4(\tau)))^2}\
 \le \ 
\frac{2}{e} \ ,
\label{astu}
\eqe
\centerline{where $\eta\,$ is defined below (\ref{eta}),}
\eq 
\quad  
\frac{ |p|\ e^{-\frac{p^2}{2\La ^2}} }{\sup(\tau |p|,\ka)}\ 
\,\le \, 
\frac{1}{\sqrt{e}}\ ,\quad  
\frac{ p^2\ e^{-\frac{p^2}{2\La ^2}} }{\sup(\tau |p|,\ka)}\ 
\,\le \,  
\ka\ \frac{2}{e}\ \ ,\quad 
\frac{ |p|^3\ e^{-\frac{p^2}{2\La ^2}}}{\sup(\tau |p|,\ka)}\ 
\,\le \,  
\ka^{2}\ (\frac{3}{e})^{3/2}\ .
\label{astx}
\eqe
\\[.1cm]
{\it Proof}~: 
a) 
The  bound  (\ref{la3}) follows from
\eq
\frac{2}{\La ^3} \
e^{-\frac{m^2}{\La ^2}}\ \le \ 
\frac{2}{\ka ^3}\ \sup_{x \ge 0}(1+x)^{3}\ e ^{-x^2}\ ,
\label{pv}
\eqe
and the function of $x\,$ is maximized for $x=\frac{\sqrt{7}-1}{2}\,$.
To prove  (\ref{p}) note
\[
p^2\
e^{-\frac{p^2}{2\La ^2}}\ \le\
{\ka}^2\ \sup_x \Bigl\{x^2\  e ^{-\frac{x^2}{2}}\Bigr\}\ =\ 
\frac{2 \ka ^2}{e}\ ,\quad
|p|\
e^{-\frac{p^2}{2\La ^2}}\ \le\
{\ka}\ \sup_{x\ge 0} \Bigl\{x\  e ^{-\frac{x^2}{2}}\Bigr\}\ =\ 
\frac{\ka}{\sqrt{e}}\ .
\]
b)
The bounds are proven similarly as in a). For $w=0$ the result follows
from a).\\
 For $|w| = 1\,,2\,,3\,$ we use (\ref{33}), (\ref{34}),(\ref{35}).
We may suppose that the axes have been chosen such that $p$ is
parallel to one of them. 
For  $|w| = 1\,$ we then find
\[ 
|\ka ^3\ \pa ^{w}\,\frac{2}{\La ^3}\,e^{-\frac{p^2}{2\La ^2}}
\,e^{-\frac{m^2}{\La ^2}}|
\]
\[
 \le
\inf\Bigl\{\frac{4}{|p|}\ \sup_{x^2} \{x^2\, e ^{-\frac{x^2}{2}}\}\,
 \sup_{y\ge 0}\{(1+y)^3\,e ^{-y^2}\}\,,\
\frac{4}{\ka}\ \sup_{x\ge 0} \{x\, e ^{-\frac{x^2}{2}}\}\,
\sup_{y\ge 0}\{(1+y)^4\,e ^{-y^2}\}\Bigr\}\ .
\]
For  $|w| = 2\,$ we obtain
\[ 
|\ka ^3\ \pa ^{w}\,\frac{2}{\La ^3}\,e^{-\frac{p^2}{2\La ^2}}
\,e^{-\frac{m^2}{\La ^2}}|
\]
\[
\le\ \inf\Bigl\{\frac{16}{|p|^2}\ \sup_{x^2} 
\{|x^4-\frac12 x^2|\, e ^{-\frac{x^2}{2}}\}
 \sup_{y\ge 0}\{(1+y)^3\,e ^{-y^2}\,,\
\frac{16}{\ka ^2}\ \sup_{x\ge 0} \{|x^2-\frac12|\, e ^{-\frac{x^2}{2}}\
\sup_{y\ge 0}\{(1+y)^5\,e ^{-y^2}\}\Bigr\}\ .
\]
For  $|w| = 3\,$ we get
\[ 
|\ka ^3\ \pa ^{w}\,\frac{2}{\La ^3}\,e^{-\frac{p^2}{2\La ^2}}
\,e^{-\frac{m^2}{\La ^2}}|
\]
\[
\le\ \inf\Bigl\{\frac{16}{|p|^3}
\ \sup_{x^2} \{|-x^6+\frac32 x^4|\, e ^{-\frac{x^2}{2}}\}
 \sup_{y\ge 0}\{(1+y)^3\,e ^{-y^2}\},\,
\]
\[
{ \qquad\qquad}
\frac{16}{\ka ^3}\ \sup_{x\ge 0} \{|-x^3+\frac32x|\, e ^{-\frac{x^2}{2}}\}\
\sup_{y\ge 0}\{(1+y)^6\,e ^{-y^2}\}\Bigr\}\ .
\]
Maximizing the expressions depending on $x$ and $y$ and taking the
maximal constant in each of the three expressions gives the numerical 
constants of  (\ref{kw}).\\
The bounds (\ref{kw1}) follow on replacing 
$\, e ^{-\frac{x^2}{2}} \to e ^{-{x^2}}\,$
in maximizing the previous expressions.\\ 
c) The first bound (\ref{astu}) follows from 
\[
\frac{1}{\sup(\ka,\eta_{1,4}^{(4)}(\tau p_1,p_2,p_3,p_4(\tau)))}\ |p_1| \ 
e^{-\frac{p_1^2}{2\La ^2}} \ \le \ 
\frac{ |p_1|}{\ka} \ 
e^{-\frac{p_1^2}{2\La ^2}} \ \le \ 
\frac{ |p_1|}{\La} \ 
e^{-\frac{p_1^2}{2\La ^2}} \ \le \ 
e^{-1/2} 
\]
and the second bound follows analogously.\\
The bounds (\ref{astx}) are obtained by the same reasoning.
\qed

\noindent
{\it Lemma 7:}\\ 
a)
\eq
\int_0^{\La} d\La' \  \La'^{-3}\ 
e ^{-m^2/\La '^2}\ \ka'^{2}\,
\log^{\la}({\ka '\over m})\ \le 
\ K_1\ \frac{\log^{\la+1}\bigr({\ka \over m}\bigl)}{\la+1}
\quad\,\, \mbox{with}\quad\, K_1\,=\,\frac{K_2}{2}\,=\, 3.1 \ ,\quad
\label{k'}
\eqe
b)
\eq
\int_0^{\La} d\La' \  \La'^{-5}\ 
e ^{-m^2/\La '^2}\ \ka'^{4}\,
\log^{\la}({\ka '\over m})\ \le 
\ K'_1\ \frac{\log^{\la+1}\bigr({\ka \over m}\bigl)}{\la+1} 
\quad \mbox{with}\quad K_1'\,=\,14.5\ \ .
\label{k1}
\eqe
\\[.1cm]
{\it Proof:}
The integrals are bounded through
\[
\int_1^{\ka/m} \frac{dx}{x}\ (\frac{x}{x-1})^s\ 
e^{-\frac{1}{(x-1)^2}} \ \log^{\la} x
\ \le \
\sup_{y \ge 0} \Bigl((1+y)^s\ e^{-y^2}\Bigr)
\frac{\log^{\la+1} (\ka/m)}{\la +1}\ ,
\]
where $s \in \{3,\,5\}\,$. The $\sup\,$ leads to the numerical constants.
\qed

\noindent
{\it Lemma 8:} For  $\la \in [0,1]$ and $x,\,y \in \mathbb{R}^d\,$, 
if $|x+y|\geq |x|$ then $|\la x+y|\geq
  \la |x|$.\\[.1cm]
{\it Proof:} 
$\ |\la x+y|\geq|x+y|-|(1-\la)x|\geq|x|-(1-\la)|x|=\la
|x|\,$. 
\qed

\section{Sharp bounds on Schwinger functions}
With the aid of the FE (\ref{feq}) it is possible 
to establish a particularly simple inductive proof of the 
renormalizability of $\vp_4^4$ theory. Renormalizability 
in fact appears  as a consequence of the
following bounds [KKS], [M\"u] on the functions
${\cal L}^{\La,\Lao}_{2n.l}\,$~:\\
\eq
\mbox{ Boundedness}\qquad
|\pa^w {\cal L}^{\La,\Lao}_{2n,l} (\vec{p})| \le\,
\ka^{4-2n-|w| }\,{{\cal P}_1}(\log {\ka  \over m})\,
{{\cal P}_2}({|\vec{p}| \over \ka})\,,\qquad\ 
\label{propo1}
\eqe
\eq
\mbox{ Convergence }\quad
|\pa_{\Lao} \pa^w {\cal L}^{\La,\Lao}_{2n,l}(\vec{p})| \le\,
{1\over \Lao^2}\, \ka^{5-2n-|w| }\,{{\cal P}_3}(\log {\Lao  \over m})\,
{{\cal P}_4}({|\vec{p}| \over \ka})\ .
\label{propo2}
\eqe
 The ${\cal P}_i$ denote 
polynomials 
with nonnegative coefficients, which 
depend on $l,n,|w|\,$,
but not on $\vec{p},\,\La,\,\ka=\La+m,\,\Lao$. 
The statement (\ref{propo2}) implies renormalizability, 
since it proves the limits
$\,\lim_{\Lao \to \infty,\;\La \to 0}  {\cal L}^{\La,\Lao}(\vec{p})\,$ 
to exist to all loop orders $l\,$.
But the statement (\ref{propo1}) has to be obtained first
to prove (\ref{propo2}). 

The standard inductive scheme which is used to prove these bounds,
and which we will also employ in the proof of the subsequent proposition,
goes up in $n+l\,$ and
for given  $n+l\,$ descends in $n\,$, and for given
$n,\,l\,$ descends in $|w|\,$.
 The r.h.s. of the FE is then prior the l.h.s. in the
inductive order, and the bounds can thus be verified 
for suitable boundary conditions on integrating the r.h.s. of the FE over
$\La\,$, using the bounds of the proposition. 
Terms with $2n+|w| \ge 5\,$ are integrated downwards from 
$\Lao$ to $\La$, since for those terms we
have the boundary conditions at $\La =\,\Lao\,$ 
following from (\ref{nawi})
\[
 \pa^w \, {\cal L}_{2n,l}^{\La,\Lao} (p_1,\ldots p_{2n-1}) 
=0\ \mbox{ for}\quad
2 n+|w| 
\ge 5\,,
\]
whereas the terms with $2n+|w| \le 4$ at the renormalization point - 
which we choose at zero momentum for simplicity -      
are integrated upwards from $0$ to $\,\La$, since they are fixed 
at $\La=\,0\,$ by
renormalization conditions, which define the relevant parameters of the
theory. 
We will choose for simplicity
\eq
 {\cal L}_{4,l}^{0,\Lao} (0,0,0) = \de_{l,0}\ \frac{g}{4!}\ ,
\quad
 {\cal L}_{2,l}^{0,\Lao} (0) = 0\ ,
\quad
\pa_{p^2} {\cal L}_{2,l}^{0,\Lao} (0) = 0\ ,
\label{ren}
\eqe
though more general choices could be accommodated for without any 
problems\footnote{It would amount to absorb the new constants 
in the respective lower bounds on $K\,$ in part B of the proof.}.

Our new result combines the sharp bounds on the high momentum 
behaviour from [KM] with good control on the constants hidden   
in the symbols $\,{\cal P}\,$ in (\ref{propo1}), (\ref{propo2}) .

In the Theorem and the Proposition we use the following notations 
and assumptions~:\\
We denote by  $\,(p_1,\ldots,p_{2n})\,$ a set of external momenta with
$p_1+\ldots+p_{2n}=0\,$, and  we define
\[
\vec{p}=(p_1,\ldots,p_{2n-1})\ ,\qquad |\vec{p}|=
\sup_{1\le i \le 2n} |p_i|\ .
\]
Furthermore   
\eq
\eta_{i,j}^{(2n)}(p_1,\ldots,p_{2n})~:=\ \inf\Bigl\{|p_i+\sum_{k\in J}
p_k|\,/\,J\subset\bigl(\{ 1,...,2 n\}-\{ i,j \}\bigr)\Bigr\}\ .
\label{eta}
\eqe 
Thus  $\eta_{i,j}^{(2n)}\,$ is the modulus of the 
smallest subsum of external momenta
containing $p_i\,$ but not $p_j\,$.
We assume $\,0\leq \La \leq \Lao\,$, and we write $\ka =\La+m\,$.

\noindent
Our main result
 can then be stated as follows~:\\

\noindent
{\bf Theorem~:}\\
There exists a constant $\,\ti K>0\,$ such that   
\\[.1cm]
\eq
  |{\cal L}_{2n,l}^{\La,\Lao}(\vec{p})|\leq \ka^{4-2n}\ 
 \frac{\ti K^{2l+n-2}}{ n!}\ 
(n+l)!\  \sum_{\la=0}^{\la =l}
\frac{\log^{\la}\bigr(\sup({|\vec{p}| \over \ka},{\ka\over
    m})\bigl)}{2^{\la}\,\la!} \quad \mbox{for}\ \ \, 2n  >  2\ , 
\label{thm1}
\eqe
\eq
  |{\cal L}_{2,l}^{\La,\Lao}(p)|\leq 
\sup(|p|,\ka)^{2}\ \frac{\ti K^{2l}}{(l+1)^2}
\ l!\,\sum_{\la=0}^{\la =l-1} 
\frac{\log^{\la}
\bigr(\sup({|p| \over \ka},{\ka\over m})\bigl)}{2^{\la}\,\la!}\ ,\quad
l \ge 1\ .
\label{thm3}
 \eqe
The Theorem follows from the subsequent Proposition.
In the Proposition the bounds are presented in a form 
such that they can  serve at the same time as an induction hypothesis
for the statements to be proven. We then have to  include also bounds on 
momentum derivatives of the Schwinger functions in order to have a 
complete inductive scheme.\\

\noindent{\bf Proposition~:}\\
 We assume $|w| \le 3\,$, where the derivatives are taken w.r.t.
some momentum $p_i\,$. Furthermore
$j \in \{1,\ldots,2n\}/ \{i\}\,$. 
There exists a constant $\,K>0\,$ such that for  
$\, 2n  > 4\,$ \\[.1cm]
\eq
  |\pa^{w} {\cal L}_{2n,l}^{\La,\Lao}(\vec{p})|\leq \ka^{4-2n}\ 
 \frac{K^{2l+n-2}}{(l+1)^2\, n!\,n^3}\ (n+l-1)! 
\ {1 \over \bigl(\sup(\ka,\eta_{i,j}^{(2n)})\bigr)^{|{w}|}}\ 
\sum_{\la=0}^{\la =l}
\frac{\log^{\la}\bigr(\sup({|\vec{p}|\over \ka},{\ka\over
    m})\bigl)}{2^{\la}\,\la!}
\ .
\label{prop1}
\eqe
For $\, 2n = 4\,,\; |w| \ge 1$
\eq
  |\pa^{w} {\cal L}_{4,l}^{\La,\Lao}(\vec{p})|\ \leq\ 
\frac{K^{2l-1/4}}{(l+1)^2\, 2^4}\ (1+l)!
\ {1 \over \bigl(\sup(\ka,\eta_{i,j}^{(4)})\bigr)^{|w|}}\ 
 \sum_{\la=0}^{\la =l-1}
\frac{\log^{\la}\bigr(\sup({|\vec{p}| \over \ka},{\ka\over
    m})\bigl)}{2^{\la}\,\la!}\bigl)
\ .
\label{prop2}
\eqe
For $\, 2n = 4\,,\; |w| =0\,$
\eq
  | {\cal L}_{4,l}^{\La,\Lao}(\vec{p})|\ \leq\ 
\frac{K^{2l}}{(l+1)^2\, 2^4}
\ (1+l)!\, \sum_{\la=0}^{\la =l-1}
\frac{\log^{\la}\bigr(\sup({|\vec{p}|\over \ka},{\ka\over
    m})\bigl)}{2^{\la}\,\la!}\ 
\Bigl(1+\log\bigr(\sup({|\vec{p}| \over \ka},\frac{\ka}{m})\bigr)\Bigr)\ ,
\label{prop3}
\eqe
\eq
  | {\cal L}_{4,l}^{\La,\Lao}(0,p_2,p_3)|\ \leq\ 
\frac{K^{2l}}{(l+1)^2\, 2^4}
\ (1+l)!\, \sum_{\la=0}^{\la =l}
\frac{\log^{\la}\bigr(\sup({|\vec{p}| \over \ka},{\ka\over
    m})\bigl)}{2^{\la}\,\la!}\ .
\label{prop30}
\eqe
For  $2n=2\,,\; |w| =3\,$
\eq
|\pa^{w} {\cal L}_{2,l}^{\Lambda,\Lambda_0}(p)|\leq
\sup(|p|,\ka)^{-1}\ \frac{K^{2l-1-\frac14}}{(l+1)^2}
 \ l!\ \sum_{\la=0}^{\la =l-2}
\frac{\log^{\la}\bigr(\sup({|p| \over \ka},{\ka\over m})\bigl)}{2^{\la}\,\la!}
\  .
\label{prop4}
 \eqe
For  $2n=2\,,\; 0\le |{w}| \le 2\,,\ l \ge 2\,$
\eq
  |\pa^{w} {\cal L}_{2,l}^{\La,\Lao}(p)|\leq 
\sup(|p|,\ka)^{2-|{w}|}\ \frac{K^{2l-1}}{(l+1)^2}
\ l!\,\sum_{\la=0}^{\la =l-2} 
\frac{\log^{\la}\bigr(\sup({|p| \over \ka},{\ka\over m})\bigl)}{2^{\la}\,\la!}
\ \Bigr(1\,+\, \log(\sup({|p| \over \ka},\frac{\ka}{m}))\Bigl)
\ .
\label{prop5}
\eqe
For  $2n=2\,,\;  |{w}|\in \{0,\,2\}\,,\ l \ge 2\,$
\eq
  |\pa^{w} {\cal L}_{2,l}^{\La,\Lao}(0)|\leq 
\ka^{2-|{w}|}\ \frac{K^{2l-1}}{(l+1)^2}
\ l!\,\sum_{\la=0}^{\la =l-1} 
\frac{\log^{\la}\bigr({\ka\over m}\bigl)}{2^{\la}\,\la!}
\ .
\label{prop6}
\eqe
{\it Remarks~:}\\
Note that $j\,$ in (\ref{prop1}) - (\ref{prop2}) is 
otherwise arbitrary apart from the condition  $j\neq i\,$,
so that the bound arrived at will be in fact 
\[
  |\pa^{w} {\cal L}_{2n,l}^{\La,\Lao}(\vec{p})|\leq 
\]
\[
\ka^{4-2n}\ \frac{K^{2l+n-2}}{(l+1)^2\, n!\,n^3}\
\inf_{j,1\le j\le 2n}   
{1 \over\bigl( \sup(\ka,\eta_{i,j}^{(2n)})\bigr)^{|w|}} 
 (n+l-1)!\sum_{\la=0}^{\la =l}
\frac{\log^{\la}\bigr(\sup({|\vec{p}|\over \ka},{\ka\over m})\bigl)}{\la!}
\ .
\]
We will choose $\,j=2n\,$ in the proof.
 This means that the momentum $p_{2n}\,$ 
will be eliminated on both sides  of the FE.\\
Since the elementary vertex has a weight 
$\frac{g}{4!}\,$, a perturbative Schwinger function
 ${\cal L}_{2n,l}\,$ carries a factor $(\frac{g}{4!})^{l+n-1}\,$
For simplicity of notation 
{\it we replace this factor by one in the subsequent
proof}. So the final
numerical bound on the Schwinger functions in terms of the constant
$K\,$, see (\ref{K}) below, should be multiplied by this factor.\\[.2cm] 
{\it Proof~:}\\
The above described inductive scheme starts from the constant
${\cal L}^{\La,\Lao}_{4,0}\,$ at loop order 0.
From this term, irrelevant tree level
terms with $\,n>2\,$ are  produced by the second
term on the r.h.s. of the FE. For those terms the Proposition is verified
from a simplified version of part A) II) of the proof,
 where all sums over loops are
suppressed. Note also that the two-point function for $l=1$ is given by the
momentum independent tadpole which is bounded by $\ka ^2\,$.
We will subsequently assume that $l \ge 1\,$ for simplicity of
notation.\\[.1cm] 
A) {\bf Irrelevant terms} with $2n+|w|\geq 5$~:\\[.1cm]
I) {\it The first term on the r.h.s. of the FE}\\[.1cm]
a)  $2n> 4\,$:\\[.1cm] 
Integrating the FE (\ref{feq}) w.r.t. the flow
parameter $\ka'$ from $\ka\,$ to $\Lao+m\,$ gives the
following bound for the  first term on the r.h.s. of the FE 
- denoting $\,\La'=\ka'-m\,$ and, as a shorthand, $|\vec{p}|_{2n+2} =\sup
(|\vec{p}|, \,|k|,\, |-k|)$ = $\sup
(|p_1|,\ldots,|p_{2n}|, \,|k|)\,$,
$\eta_{i,2n}^{(2n+2)}=\eta_{i,2n}^{(2n+2)}(\vec  p, k,-k)\,$
 ~:
\[
\frac{(2n+1)(2n+2)}{2}
\int_\ka^{\Lao+m} d\ka' \int_k {2 \over \La'^3}\
e^{-{{k^2+m^2}\over \La'^2}}\
 \ka'^{4-(2n+2)}\ \frac{K^{2l+n-3}}{l^2\, (n+1)!\,(n+1)^3}
\]
\[
\times\, (n+l-1)!\  {1 \over
      \bigl({\sup(\ka',\eta_{i,2n}^{(2n+2)})\bigl)^{|w|}}}
\sum_{\la=0}^{\la =l-1}  
\frac{\log^{\la}(\sup({|\vec{p}|_{2n+2} \over \ka'},{\ka'\over m}))}
{2^{\la}\,\la!}
\]
\eq
\leq \  (\frac{n}{n+1})^3\ (2n+1)\ \frac{K^{2l+n-3}}{l^2\, n!\,n^3} \
(n+l-1)!\ \sum_{\la=0}^{\la =l-1}\frac{1}{2^{\la}\,\la!}
\label{1stin}
\eqe
\[
\times\ 
K_2\, \int_\ka^{\Lao+m} d\ka'\ \ka'^{3-2n-|w|} \ 
\int_k \ \frac{1}{\ka'^4}\
{1\over     \bigl({\sup(1,{\eta_{i,2n}^{(2n+2)}\over  \ka'})}\bigr)^{|w|}}
\ e^{-{k^2 \over {\La'^2}}}\
\log^{\la}\bigr(\sup({|\vec{p}|_{2n+2} \over \ka'},{\ka'\over m})
\bigl)\ .
\]
We used Lemma 6, (\ref{la3}).  
We bound the momentum integral as follows,
setting $\,x =\frac{k}{\ka '}\,$:
\[
\int_x 
{1\over     \bigl({\sup(1,{\eta_{i,2n}^{(2n+2)}\over  \ka'})}\bigr)^{|w|}}
\ e^{-{x^2}}\
\log^{\la}\bigr(\sup({|\vec{p}|_{2n+2} \over \ka'},{\ka'\over m})\bigl) \ \le \
\]
\eq
\sup_x \Bigl\{ e^{-\frac{x^2}{2}}\ 
{1\over     \bigl({\sup(1,{\eta_{i,2n}^{(2n+2)}\over
      \ka'})}\bigr)^{|w|}}\Bigr \}\
\int_x e^{-\frac{x^2}{2}}\  
\log^{\la}\bigr(\sup({|\vec{p}|_{2n+2} \over \ka'},{\ka'\over m})\bigl)
\ . 
\label{sep}
\eqe
The first term is bounded\footnote{by the definition of $\eta$ (\ref{eta}) we
have $\,\eta_{i,2n}^{(2n+2)} \in \{ |q|, |q\pm k|\}\,$, if
$\,\eta_{i,2n}^{(2n)} = |q|\,$.}
 with the aid of Lemma 3, (\ref{ineq1}), as
\[
\sup_x \Bigl\{ e^{-\frac{x^2}{2}}\ 
{1\over     
\bigl({\sup(1,{\eta_{i,2n}^{(2n+2)}\over  \ka'})}\bigr)^{|w|}}\Bigr \}
\  \le \ c(|w|)\
{1\over     
\bigl({\sup(1,{\eta_{i,2n}^{(2n)}\over  \ka'})}\bigr)^{|w|}}
\ .
\]
To bound the  integral in (\ref{sep}),
we note that
\[
\sup({|\vec{p}|_{2n+2} \over \ka'},{\ka'\over m})\ \le \
\sup(\frac{|\vec{p}|}{ \ka'}+\frac{|k|}{ \ka'},
\frac{\ka'}{m})
\]
so that the integral can be bounded using  
\eq
\int_{x}   e^{-\frac{x^2}{2}}\ \log^{\la}(\sup(|x|+a,b))\ \le \
\int_{x}   e^{-\frac{x^2}{2}}\ \log^{\la}(|x|+a)\ + \
\int_{x }   e^{-\frac{x^2}{2}}\ \log^{\la}b
\eqe
with $\,a=\frac{|\vec{p}|}{ \ka'}\,$ and $\,b=\frac{\ka'}{m}\,$.
We have
\eq
\int_{x }   e^{-\frac{x^2}{2}}\ \log^{\la}b\ =\ 
\frac{1}{4 \pi ^2} \  \log^{\la}b\ .
\label{bint}
\eqe
Using Lemma 4, (\ref{ineq6}) and $\,\frac{1}{4\pi ^2}+\frac14 \le
\frac13\,$, we can then bound the
integral from (\ref{sep}) by 
\eq
\int_{x}  e^{-\frac12 x^2}\ 
\log^{\la}(\sup({|\vec{p}|_{2n+2} \over \ka'},{\ka'\over m}))\ \le \
K_3 \ \Bigl(\,\log^{\la}(\sup({|\vec{p}| \over \ka'},{\ka'\over m}))
\,+\,   [\la!]^{1/2}\,\Bigr)\ ,
\label{momint}
\eqe
where 
\eq
K_3 \ =\ \frac13 \  .
\label{K2}
\eqe 
With these results (\ref{1stin}) can now be bounded by
\eq 
 (\frac{n}{n+1})^3\ 
(2n+1)\ \frac{K^{2l+n-3}}{l^2\, n!\, n^3}\ (n+l-1)! \ K_2  \ K_3
{c(|w|)\over     \bigl({\sup(1,{\eta_{i,2n}^{(2n)}\over  \ka})}\bigr)^{|w|}}
\label{11}
\eqe
\[
\times\ \sum_{\la=0}^{\la =l-1} 
\int_\ka^{\Lao+m} d\ka'\ \ka'^{3-2n-|w|}\ 
\frac{1}{2^{\la}\,\la!}
\Bigl(\log^{\la}(\sup({|\vec{p}| \over \ka'},{\ka'\over m}))
\,+\,  [\la!]^{1/2}\,\Bigr)\ .
\]
Using Lemma 5,  (\ref{weh1})
we find - writing $s=2n+|w|-4\,$ -
\[
\sum_{\la=0}^{\la =l-1}\frac{1}{2^{\la}\,\la!}\ 
\int_\ka^{\Lao+m} d\ka'\ \ka'^{-s-1} \ 
\Bigl(\,\log^{\la}(\sup({|\vec{p}|_{2n} \over \ka'},{\ka'\over m}))
\,+\,   [\la!]^{1/2}\,\Bigr)
\]
\[
 \ \le\ \frac{\ka^{-s}}{s}\Bigl\{
\ 
3\ \sum_{\la=0}^{\la =l-1} \frac{1}{2^{\la}\,\la!}\,
\log^{\la}\sup({|\vec{p}|\over \ka},{\ka \over m})\,+\,2\
\Bigr\}
\le\
5\ \frac{\ka^{-s}}{s}\sum_{\la=0}^{\la =l-1} \frac{1}{2^{\la}\,\la!}\,
\log^{\la}\sup({|\vec{p}| \over \ka},{\ka \over m})\ .
\]
Using this bounds in (\ref{11}), the first term on the
r.h.s. of the FE  then satisfies the 
induction hypothesis (\ref{prop1})\footnote{we may note
  that for this term the sum extends up to $l-1$ only}, 
\[ 
\ka^{4-2n} 
\ \frac{K^{2l+n-2}}{(l+1)^2\, n!\, n^3} \ (n+l-1)!\ 
{1 \over     \bigl({\sup(\ka,\eta_{i,2n}^{(2n)})}\bigr)^{|w|}}
\sum_{\la=0}^{\la =l-1}\frac{1}{2^{\la}\,\la!}\,
\log^{\la}\sup({|\vec{p}| \over \ka},{\ka \over m})\ ,
\]
on imposing the lower bound on $K\,$
\eq
K^{-1} \, (\frac{n}{n+1})^3\ (2n+1)\, \frac{(l+1)^2}{l^2}\ 
K_2\ K_3\  c(|w|)\
\frac{5}{(2n +|w|-4)}\  \le\ 1 \ .
\label{bdk0}
\eqe

\noindent
b)  $2n= 4\,,\ |w| \ge 1$:\\ 
The only change w.r.t. part a) is that we have to verify the bound
with an addditional factor of $K^{-1/4}\,$  appearing in
(\ref{prop2}). We therefore arrive at the bound 
\eq
K^{-\frac34}\  (\frac{2}{3})^3\ 5\   
\frac{(l+1)^2}{l^2}\  K_2\  K_3\ c(|w|)\,
\frac{5}{|w|}\ \le\ 1 \ .
\label{bdk1}
\eqe

\noindent
c) $\,2n =2\,,\ |w| =3\,$:\\
Due to the momentum derivatives the corresponding contribution for
$l=1\,$ vanishes.
Using the induction hypothesis on 
$\,|\pa^w {\cal L}_{4,l-1}^{\La,\Lao}(\vec{p})|\,$ for $ l\ge 2\,$
as in (\ref{1stin})  we obtain in close analogy with A) I) a) and b)
the following bound 
\[
 (\frac{1}{2})^3\  \frac{K^{2l-1-\frac14}}{(l+1)^2}\
\frac{\ka^{2}}{\sup(|p|,\,\ka)^3}\
l! \sum_{\la=0}^{\la =l-2}
\frac{\log^{\la}\bigr(\sup({|p|\over \ka},{\ka\over m})\bigl)}{2^{\la}\,\la!}
\]
in agreement with (\ref{prop4}),
on imposing the lower bound
\eq
K^{-1}\ \frac38\ \frac{(l+1)^2}{l^2} \  K_2\  K_3\ c(3)\ 5\
\le \ 1\ .
\label{bdk2}
\eqe
\\

\noindent
II) {\it The second term on the r.h.s. of the FE}\\[.1cm]
a) $2n >4\,$:\\[.1cm] 
We sum over all contributions without taking into account the fact
that some of them are suppressed by
supplementary fractional powers of $\,K\,$.
Some additional precaution is required in the presence of relevant
terms, i.e. underived   four-point functions, and two-point functions
derived at most twice. These functions are decomposed as
\eq  
 {\cal L}_{4,l}(p_1,p_2,p_3)\,=\,
 {\cal L}_{4,l}(0,p_2,p_3)\,+\,
p_{1,\mu}\int_0^1 d\tau\ \pa_{1,\mu}\,{\cal L}_{4,l}(\tau
p_1,p_2,p_3)\ .
\label{ast}
\eqe
For the two-point function we may suppose without limitation that 
$p=(p_0,0,0,0)\,$. We then write $p$ instead of $p_0\,$,
$\pa$ instead $\frac{\pa}{\pa p}\,$ and interpolate 
\eq
\pa^{2} {\cal L}_{2,l}(p)=\,
\pa^{2} {\cal L}_{2,l}(0)\,+\,
p \int_0^1 d\tau \
\pa^3 {\cal L}_{2,l}(\tau p) \ ,
\label{sl2}
\eqe
\eq
\pa {\cal L}_{2,l}(p)=\,
p \,\pa^2 {\cal L}_{2,l}(0) \,+\,
p^2 \int_0^1 d\tau \ (1-\tau)\ \pa^3
{\cal L}_{2,l}(\tau p) \ ,
\label{sl1}
\eqe
\eq
{\cal L}_{2,l}(p)=\,
{\cal L}_{2,l}(0)\,+\,
\frac12\ 
p^2 \,\pa^2 {\cal L}_{2,l}(0) \,+\,
p^3 \int_0^1 d\tau\ \frac{(1-\tau)^2}{2!}\ \pa^3
{\cal L}_{2,l}(\tau p) \ .
\label{sl0}
\eqe
In case of the four-point function we 
use the bound from (\ref{prop30})
for the first term of the decomposition, 
and the bound from (\ref{prop2}) for the second term. 
Here the 
interpolated momentum $p_1\,$ 
will be (without loss of generality) supposed
to be the momentum $q\,$ of the propagator 
linking the two terms on the r.h.s. of the FE. We then will use the
bound (\ref{astu}) to get rid of the momentum factor produced through
interpolation. {\it Thus we can avoid using 
(\ref{prop3}) which would not reproduce a bound matching with our
induction hypothesis.}
For the two-point function we similarly use either the bounds (\ref{prop6})
at zero momentum, or  (\ref{prop4}), 
together with (\ref{astx}) and (\ref{la3}), for the interpolated term.\\
These decompositions lead to additional factors
in the bounds. So as not to produce too lengthy expressions
we will first write the bounds only for the contributions where
the additonal factors are not present and add the modifications
necessitated by those terms afterwards
(see after (\ref{bdk6})).\\
A second point has to be clarified (which is treated in a fully
explicit though notationally more complex way in [GK]).
When deriving both sides of the flow equation w.r.t. the momentum
$p_i$, there may arise two situations for the second term on the
r.h.s. :
either the two momenta $p_i\,$ and $p_{2n}\,$ appear 
both as external momenta of only one term 
$\,{\cal L}_{n_i,l_i}\,$,  or each of them appears in a different
  $\,{\cal L}_{n_i,l_i}\,$.
In the first case the derivatives only apply to the term
where they both appear, 
and not to the second one which is independent of $p_i\,$,
nor to the propagator  linking the two terms.
In the second case also the other term and the linking propagator
depend on $p_i$ via the momentum $q\,$ of the propagator 
which is a subsum of momenta containing $p_i\,$.
Applying then the induction hypothesis to both terms
we get a product of $\eta$-terms which can be bounded 
by a single one~:
\eq
{1 \over \bigl(\sup(\ka,\eta_{i,2n_1}^{(2n_1)})\bigr)^{|w_{1}|}}\ 
{1 \over \bigl(\sup(\ka,\eta_{i,2n}^{(2n_2)})\bigr)^{|w_{2}|}}
\ \le \
{1 \over \bigl(\sup(\ka,\eta_{i,2n}^{(2n)})\bigr)^{|w_{1}|
+|w_{2}|}}\ ,
\label{eta1}
\eqe
since one verifies that the set of momenta over which the
$\inf$ is taken in $\,\eta\,$ in the terms on the l.h.s. of (\ref{eta1}) is 
contained in the one on the r.h.s. of (\ref{eta1}). 
Here $\eta_{i,2n_1}^{(2n_1)}\,$ has been introduced as in (\ref{eta})
for the momentum set $\,\{p_{1}, \ldots, p_{2n_1-1},q\}\,$,
where $\,q=-p_1-p_2-\ldots-p_{2n_1-1}\,$,
and  we understand (without introducing new  notation)
that $\,\eta_{i,2n}^{(2n_2)}\,$ has been introduced as in (\ref{eta})
for the momentum set $\,\{ q, p_{2n_1}, \ldots, p_{2n}\}\,$
where $q\,$ takes the role of $p_i\,$.
The reasoning remains the same, if permutations 
of these momentum sets are considered, which still leave 
$p_{i}\,$ and $p_{2n}\,$ in different sets. \\
Integrating the inductive bound on the 
{\sl second} term on the r.h.s. of the FE from  $\ka\,$ to $\Lao+m\,$  
then gives us  the following bound - where we also understand that the
$\sup$ w.r.t. the previously mentioned permutations has been taken
for the momentum attributions
\[
\int_\ka^{\Lao+m} d\ka' \  \ka'^{8-(2n+2)}\ K^{2l+n-3}\ 
\sum_{\begin{array}{c}_{l_1+l_2=l},\\[-.2cm]
_{w_1+w_2+w_3=w,}\\[-.2cm]
_{n_1+n_2=n+1} \end{array} } 2\,c_{\{w_i\}}\  
\frac{ n_1}{(l_1+1)^2\, n_1!\,n_1^3} \ \frac{ n_2}{(l_2+1)^2\, n_2!\,n_2^3} 
\]
\[
\times\
{1 \over \bigl(\sup(\ka',\eta_{i,2n_1}^{(2n_1)})\bigr)^{|w_{1}|}}\ 
(n_1+l_1-1)!\sum_{\la_1=0}^{\la_1 =l_1} 
\frac{\log^{\la_1}\bigr(\sup({|\vec{p}| 
\over \ka '},{\ka '\over m})\bigl)}{2^{\la_1}\,\la_1!}\
\frac{2}{\La'^3}\ |\pa^{w_3} \  e ^{-\frac{q^2+m^2}{\La '^2}}|
\]
\[
\times\
{1 \over \bigl(\sup(\ka',\eta_{i,j_2}^{(2n)})\bigr)^{|w_{2}|}}\ 
(n_2+l_2-1)!\ \sum_{\la_2=0}^{\la_2 =l_2}
\frac{\log^{\la_2}
\bigr(\sup({|\vec{p}|\over \ka '},{\ka'\over m})\bigl))}
{2^{\la_2}\,\la_2!}\ . 
\]
We use (\ref{eta1}) to bound the previous expression by
\[
\sum_{\begin{array}{c}_{l_1+l_2=l},\\[-.2cm]
_{n_1+n_2=n+1},\\[-.15cm]_{\la_1 \le l_1, \,\la_2 \le l_2}
 \end{array} } 
\frac{1}{(l_1+1)^2\, (l_2+1)^2}\ \frac{1}{n_1^2\,n_2^2} 
\ \frac{n!}{ n_1!\,n_2!} \ \frac{(\la_1+\la_2)!}{\la_1!\,\la_2!} 
\ \frac{(n_1+l_1-1)!\ (n_2+l_2-1)!}{(n+l-1)!} 
\]
\[
\times\ 2\ K^{2l+n-3}\ \frac{(n+l-1)!}{n!} 
 \int_\ka^{\Lao+m} d\ka' \  \ka'^{3-2n}\
\frac{\log^{\la_1+\la_2}
\bigr(\sup({|\vec{p}|\over \ka '},{\ka '\over m})\bigl)}
{2^{\la_1+\la_2}\,(\la_1+\la_2)!}
\]
\[
\times
\sum_{w_1+w_2+w_3=w}
 c_{\{w_i\}}\ \frac{2}{\La'^3}\  |\pa^{w_3} \ e ^{-\frac{q^2+m^2}{\La '^2}}|\
{1 \over \bigl(\sup(\ka',\eta_{i,2n}^{(2n)})\bigr)^{|w_{1}|+|w_{2}|}}\ .
\]
 Using Lemma 2, (\ref{le5-1}) and Lemma 6, (\ref{kw1}),
and the fact that
\[
\sup(|q|, \ka')^{-|w_3|}\
{1 \over \bigl(\sup(\ka',\eta_{i,2n}^{(2n)})\bigr)^{|w_{1}|+|w_{2}|}}
\ \le\
{1 \over \bigl(\sup(\ka',\eta_{i,2n}^{(2n)})\bigr)^{|w|}}
\]
we then arrive at the bound 
\[
K_0\ \frac{1}{(l+1)^2}\ \frac{1}{n^2}\ 2\ K^{2l+n-3}\  
\frac{1}{n!}\ (n+l-1)!\int_\ka^{\Lao+m} d\ka' \  \ka'^{3-2n-|w|}\
\sum_{0\le \la \le l} \frac{\log^{\la}
\bigr(\sup({|\vec{p}|\over \ka '},{\ka '\over m})\bigl)}{2^{\la}\,\la!}
\]
\eq
\times
\sum_{w_i}
 c_{\{w_i\}}\, K^{(|w_3|)}\
{1 \over \bigl(\sup(1,\frac{\eta_{i,2n}^{(2n)}}{\ka'})\bigr)^{|w|}}\ .
\label{2ir}
\eqe
Using also Lemma 5 we verify the bound (\ref{prop1})
\[
\ka^{4-2n}\ K^{2l+n-2}\ 
\frac{1}{(l+1)^2}\ \frac{1}{n^3} 
\frac{1}{n!}\ (n+l-1)!
\sum_{0\le \la \le l} \frac{\log^{\la}
\bigr(\sup({|\vec{p}| \over \ka },{\ka \over m})\bigl)}{2^{\la}\,\la!}
\ {1 \over \bigl(\sup(\ka,\eta_{i,j}^{(2n)})\bigr)^{|w|}}\ ,
\]
on imposing the lower bound on $K\,$
\eq
K^{-1}\ 3\cdot 2\ K_2\  \frac{n}{2n+|w|-4}\ K_0 
\sum_{w_i} c_{\{w_i\}}\, K^{(|w_3|)}\
\le \ 1\ ,
\quad n >2\ .
\label{bdk3}
\eqe
\\[.1cm]
\noindent
b) $2n=4, |w| \ge 1\,$~:\\[.1cm]
We obtain in the same way, using Lemma 2c)
\eq
K^{-3/4}\ 6\ K_2\  2\ K_0''   \sum_{{\{w_i\}}}
 c_{\{w_i\}}\,  K^{(|w_3|)}\
 \le 1 \ .
\label{bdk34}
\eqe
\\[.1cm]
\noindent
c)  $2n=2, |w| = 3\,$:\\[.1cm]
For the two-point function we obtain 
\eq
K^{-3/4}\ 6\ K_2\  K_0'' \sum_{{\{w_i\}}}
 c_{\{w_i\}}\, K^{(|w_3|)}\
\le 1 \ .
\label{bdk32}
\eqe
\\
\noindent
 {\it Taking both contributions from
the r.h.s. of the FE together}, the lower bounds on $K\,$ become  
for $n >2\,$
\eq
K_2\ \Bigl( 5\ K_3\ (\frac{n}{n+1})^3\
\frac{c(|w|)\,(2n+1)\,(l+1)^2}{(2n +|w|-4)
\,l^2} \ +\
\frac{6\ n}{2n+|w|-4} \  K_0\sum_{{\{w_i\}}}
  c_{\{w_i\}}\, K^{(|w_3|)}\Bigr)\  \le K\ ,
\label{bdk4}
\eqe
and for  $\,n =2\,$ resp. $\,n =1$ 
\eq
K_2\  \Bigl( 5\cdot 5\ (\frac{2}{3})^3\ K_3\  c(|w|)\,  
\frac{l+1)^2}{|w|\,l^2} \ +\  
 6\cdot 2\  K_0''\ \frac{2}{|w|} \sum_{\{w_i\}}
 c_{\{w_i\}}\, K^{(|w_3|)} 
\Bigr)\ \le K^{\frac34} \ , 
\label{bdk5}
\eqe
\eq
K_2\  \Bigl( 5\cdot \frac38 \ K_3\ c(3)\ 
\frac{(l+1)^2}{l^2}\  K^{-\frac14} \ +\  
6\ K_0''  \sum_{\{w_i\}}
 c_{\{w_i\}} \, K^{(|w_3|)} 
\Bigr)\ \le K^{\frac34} \ .
\label{bdk6}
\eqe
\\[.1cm]
We now come back to the modifications required because of the
decompositions (\ref{ast}), (\ref{sl2}), (\ref{sl1}), (\ref{sl0}).
We introudce the shorthands $\sum_{\{w_i\}} c_{\{w_i\}} \, K^{(|w_3|)}\equiv
\ti K(w)\equiv
\ti K\,$ and $\sum_{\{w_i\}} c_{\{w_i\}} \, K'^{(|w_3|)}\equiv
\ti K'(w)\equiv\ti K'\,$.
In order not to inflate too much  the values of the constants we distinguish
different cases. In each case we have to replace the factors
$\, K_0\,\ti K\,$ from (\ref{bdk3}) resp. 
$\, K_0''\,\ti K\,$ 
from (\ref{bdk34}) and from  (\ref{bdk32}) by the following ones:
\\
i) $n >3$~: 
\[
\frac{K_0}{2}\, \ti K\,\,+\,
2 K_0'\, \ti K\,+\,2 K_0'\, \frac{2}{\sqrt{e}\, K^{1/4}}\, \ti K'
\,+\,
2 K_0''\, \ti K\,+\, K_0''\, (\frac{1}{\sqrt{e}}
\,+\,  \frac12\ \frac{2}{e}\,+\, \frac{1}{K^{1/4}})\, \ti K'\,\ ,
\]
ii) $n =3$~: 
\[
\frac{K_0}{2}\, \ti K\,+\, 
K_0'\,\ti K\,+\, K_0'\,(\frac{2}{\sqrt{e}\, K^{1/4}}
\,+\,\frac{2}{e\,K^{1/2}})\, \ti K'\,+\, 
2 K_0''\, \ti K\,+\,2 K_0''\, (\frac{1}{\sqrt{e}}\,+\, \frac12\ 
\frac{2}{e}\,+\, \frac{1}{K^{1/4}})\, \ti K'\,,
\]
iii) $n =2$~: 
\[
2\ K_0''\, \ti K\,+\, 2\ K_0''\, \frac{1}{\sqrt{e}\, K^{1/4}}2\
\, \ti K'\ ,
\]
iv) $n =1$~: 
\[
K_0''\, \ti K\,+\,K_0''\,  (\frac{1}{\sqrt{e}}\,+\, \frac12\ 
\frac{2}{e}\,+\, \frac{1}{K^{1/4}})\, \ti K'\ .
\]
These factors can be understood as follows~:\\
In case i) we may replace $\, K_0\,$ by  
 $\, K_0/2\,$ if no two- or four-point functions appear
by Lemma 2, (\ref{le5-11}). In the other cases we use 
Lemma 2, (\ref{le5-2}) or(\ref{le5-3}), and we use the 
decompositions which then give rise to a sum of contributions.
Factors of $2\,$ appear if there exist two contributions
of the required type.
To bound the individual terms from the decomposition we also
have to use Lemma 6 c), since there appear momentum dependent factors
in the interpolation formulas which have to be bounded with the aid 
of the regularizing exponential. 
The terms  multiplied by $\ti K\,$ thus arise from the boundary
terms, those   multiplied by $\ti K'\,$ from interpolated ones
where the bounds  (\ref{kw}) instead of (\ref{kw1}) have to be used
since  the regularizing exponential has to be split up and used
for bounding two types of momentum factors.
In the cases  $n =2$ and $n =1$  there appear one four- and on
two-point function resp. two two-point functions on the r.h.s. of the
FE. Only one of these factors has to be decomposed however, since 
in the final bound we can tolerate {\it one} factor of
$\,(1+\log\bigr(\sup({|\vec{p}| \over \ka},\frac{\ka}{m})\bigr)\,$
according to the induction hypotheses for these two cases,
see (\ref{prop3}), (\ref{prop5}).

The final lower bound on $K\,$ which also turns out to be the most
stringent one in the end, stems from the case $n = 3\,$.
It is thus the following one 
\[ 
\Bigl\{ 5 K_3\ (\frac{3}{4})^3\ 
\frac{c(|w|)\,7\,(l+1)^2}{l^2}\,+\,18\
\Bigl[\frac{K_0}{2}\, \ti K\,+\, 
K_0'\,\ti K\,+\, K_0'\,(\frac{2}{\sqrt{e}\, K^{1/4}}
\,+\,\frac{2}{e\,K^{1/2}})\, \ti K'
\]
\eq
\,+\, 
2 K_0''\, \ti K\,+\,2 K_0''\, (\frac{1}{\sqrt{e}}\,+\, \frac12\ 
\frac{2}{e}\,+\, \frac{1}{K^{1/4}})\, \ti K'\Bigr]
\Bigr\}\ \frac{K_2}{2+|w|} \ \le\ K\ .
\label{bdke}
\eqe
The numerical lower bound on $\,K\,$ deduced from 
(\ref{bdke}) in the worst case $|w|=3\,$ is
\eq
K \ \ge \ 6.2 \cdot 10^5 \ .
\label{K}
\eqe
One could certainly gain several orders of magnitude by 
more carefully bounding individual special cases (see above
for one point). {\it The basic source of the (still) large numerical 
constant is  in the fact that we have to reconstruct the relevant 
terms from their derivatives.}\\

\noindent
B) {\bf Relevant terms} with $\,2n+|w| \le 4\,$:\\[.1cm] 
a) $\,2n=4\,,\ |w|=0\, $:\\[.1cm]  
 We first look at 
$\, {\cal L}_{4,l}^{\La,\Lao}(\vec{0})\,$ which is decomposed as
\eq
 {\cal L}_{4,l}^{\La,\Lao}(\vec{0})\,=\,
 {\cal L}_{4,l}^{0,\Lao}(\vec{0})\,+\,
\int_0^{\La} d\La'\ \pa_{\La'}  {\cal L}_{4,l}^{\La',\Lao}(\vec{0})\ ,
\label{4ptrel}
\eqe
where the first term is vanishes for $l \ge 1\,$, see (\ref{ren}). 
For the second term
we obtain  by induction from the { \it first} term on the r.h.s. of the FE
the bound
\[
({6 \atop 2}) \
\int_m^{\La+m} d\ka' \int_k \frac{2}{\La'^3}\ e^{-{{k^2+m^2}\over \La'^2}}
\, \ka'^{-2}\ \frac{K^{2l-1}}{l^2\, 2\cdot 3^4}\ 
(1+l)!\sum_{\la=0}^{\la =l-1}  
\frac{\log^{\la}(\sup({|k| \over \ka'},{\ka'\over m}))}{2^{\la}\,\la!}
\]
\eq
\leq \  K_2\ K_3\ ({6 \atop 2})\ \frac{1}{2\cdot 3^4}\
\frac{K^{2l-1}}{l^2} \
(1+l)! \sum_{\la=0}^{\la =l-1}
\frac{1}{2^{\la}\,\la!}\
\int_m^{\La+m} d\ka'\ \ka'^{-1} \ 
\Bigl(\,\log^{\la}({\ka'\over m})
\,+\,   (\la!)^{1/2}\,\Bigr)\ ,
\label{1stinr}
\eqe
where we used again (\ref{la3}) and (\ref{momint}), remembering that
$\,|\vec{p}|=|k|\,$ in the present case. 
We have
\eq
\int_m^{\ka} \frac{d\ka'}{\ka'} \ 
\Bigl(\,\log^{\la}({\ka'\over m})\,+\,   
[\la!]^{1/2}\,\Bigr) \,=\,
\frac{\log^{\la+1}(\frac{\ka}{m})}{\la+1} \,+\,   
\log({\ka\over m})\,[\la!]^{1/2}\ ,
\label{wehre}
\eqe
\eq
\sum_{\la=0}^{\la =l-1}
\Bigl(\frac{\log^{\la+1}(\frac{\ka}{m})}{2^{\la}\,(\la+1)!} \,+\,   
\log({\ka\over m})\,\frac{1}{2^{\la}\,\la!^{1/2}}\Bigr)\, \le\, 
\inf\Bigl\{ 6 \sum_{\la=1}^{\la =l}
\frac{\log^{\la}(\frac{\ka}{m})}{2^{\la}\,\la!}\ ,\ 
2 \sum_{\la=0}^{\la =l-1}
\frac{\log^{\la}(\frac{\ka}{m})}{2^{\la}\,\la!}
(1+\log\frac{\ka}{m})\Bigr\}
\  .
\label{57}
\eqe
Using the first of these bounds in (\ref{1stinr}), the first term on the
r.h.s. of the FE is  bounded
in agreement with the induction hypothesis by
\eq
\frac{K^{2l}}{(l+1)^2\, 2^4}
\ (1+l)!\, \sum_{\la=0}^{\la =l}
\frac{\log^{\la}({\ka\over  m})}{2^{\la}\,\la!}\ ,
\label{hyp}
\eqe
assuming the lower bound on $K\,$
\eq
K^{-1} \ 6\   K_2\ K_3\ ({6 \atop 2})\ \frac{2^4}{2\cdot 3^4}\
\frac{\,(l+1)^2 }{l^2} \ \le  \ 1  \   . 
\label{bdk8}
\eqe
In the contribution from the {\it second} term
on the r.h.s. of the FE we have one contribution with 
$n_1=2$ and one contribution with $n_2=1$ or vice versa. 
Integrating the FE (\ref{feq}) w.r.t. the flow
parameter at vanishing
momentum gives the  inductive  bound, using  (\ref{prop30}),
(\ref{prop6}) and Lemma 2 c)

\[
2\cdot 4\!\!\int_m^{\La+m}\!\!\! d\ka' \,  \frac{2}{\La'^{3}}
\  e ^{-\frac{m^2}{\La '^2}}\ \ka'^{2}\
K^{2l-1}\!\!\!\!\!\!  
\sum_{\begin{array}{c}_{l_1+l_2=l},\\[-.2cm]_{l_2 \ge 1}\end{array} }
\!\!\!\!\!\!    
\frac{ (1+l_1)!}{(l_1+1)^2\, 2^4} \ \frac{l_2 !}{(l_2+1)^2\, } 
\sum_{\la_1=0}^{\la_1 =l_1}
\frac{\log^{\la_1}({\ka '\over m})}{2^{\la_1}\,\la_1!}\
\sum_{\la_2=0}^{\la =l_2-1} 
\frac{\log^{\la_2}({\ka'\over m})}
{2^{\la_2}\,\la_2!}
\]
\[
\le\
\frac{16\,K^{2l-1}\,  K''_0}{(l+1)^2\, 2^4}\  (1+l)!
\int_0^{\La} \frac{d\La'}{\La'^{3}}\ 
e ^{-\frac{m^2}{\La '^2}}\ \ka'^{2} 
\sum_{\la=0}^{\la =l-1}
\frac{\log^{\la}({\ka '\over m})}{2^{\la}\,\la!}\ .
\]
With the aid of Lemma 7 a) the previous expression can be bounded 
as in (\ref{hyp}) assuming
\[
16\ K_0''\ K_1\ \le \ K\ .
\label{bdk9}
\]

\noindent
To go away from the renormalization point 
we proceed as in [KM].
In fact, we will distinguish four  different
situations as regards the momentum configurations.
The  bounds  established in part A) for the case $n=4,|w|=1\,$ 
are in terms of the functions $\eta_{i,j}^{(4)}\,$ from
(\ref{eta}). Assuming  (without loss of generality) 
\[
|p_4| \ge |p_1|\,,\ |p_2|\,,\ |p_3|\ ,
\]
we realize that $\eta_{i,4}^{(4)}\,$ is always given
by a sum of {\it at most two} momenta from the
set $\{p_1\,,\ p_2\,,\ p_3\}\,$. It is then obvious that the
subsequent cases ii) and iv) cover all possible situations.
The cases i) and iii) correspond to exceptional configurations
for which the bound has to be established before proceeding to the
general ones. The four cases are\\
i) $\{p_1\,,\ p_2\,,\ p_3\} =\{0\,,\ q\,,\ v\}$\\ 
ii) $\{p_1\,,\ p_2\,,\ p_3\}$ such that $\inf_i
\eta_{i,4}^{(4)}\,=\,\inf_i  |p_i|$ \\
iii) $\{p_1\,,\ p_2\,,\ p_3\} =\{p\,,-p \,,\ v\}$\\
iv) $\{p_1\,,\ p_2\,,\ p_3\}$ such that $\inf_i \eta_{i,4}^{(4)}\,=\,
\inf_{j\neq k}|p_j+p_k|\,$.\\[.1cm]
i) To prove the proposition in this case, i. e. (\ref{prop30}), we bound
\[
|{\cal L}^{\La,\Lao}_{4,l}(0,q,v)|\leq
\]
\[
|{\cal L}^{\La,\Lao}_{4,l}(0,0,0)|+\sum_{\mu}\int_0^1d\tau\,\Bigl(
|q_{\mu}\pa_{q_{\mu}}{\cal L}^{\La,\Lao}_{4,l}
(0,\tau q,\tau v)|+\,|v_{\mu} \pa_{v_{\mu}}{\cal L}^{\La,\Lao}_{4,l}
(0,\tau q,\tau v)|\Bigr)\ .
\]
The second term is bounded using the induction hypothesis: 
\eq
\frac{K^{2l-\frac14}}{(l+1)^2\, 2^4}\
\sum_{i=2,3}|p_i|
\int_0^1d\tau\,{ 1\over  \sup(\ka, \eta_{i,4}^{(4)}(\tau)\,)}\
(1+l)!\sum_{\la=0}^{\la =l-1}\frac{1}{2^{\la}\,\la!}\ 
\log^{\la}\bigr(\sup({|\vec{p}^{\;\tau}| \over \ka},{\ka\over
  m})\bigl)\ .
\label{2term}
\eqe
We have written $\eta(\tau)\,$ for the $\eta$-parameter in terms of the
scaled variables $\,p_2^{\tau}=\tau q\, ,\ p_3^{\tau}=\tau v\,$
and  $\vec{p}^{\;\tau}\,$ for the momentum set 
$\,(0,\,p_2^{\tau},\,p_3^{\tau})\,$. Using Lemma 8 we
find $\, \eta_{2,4}^{(4)}(\tau)=\tau |q|\,$, 
$\,\eta_{3,4}^{(4)}(\tau)=\tau |v|\,$,
and we thus obtain the following bound for (\ref{2term})
- apart from the prefactor
\eq
|q|\,\Bigl(
\int_0^{\inf(1,{\ka \over |q|})}
\!{d\tau  \over \ka}\,+\,
\int_{\inf(1,{\ka \over |q|})}^1
\!{d\tau \over \tau |q|}\, \Bigr)
\log^{\la}\bigr(\sup({|\vec{p}^{\;\tau}| \over \ka},{\ka\over
  m})\bigl)\,+\,
\biggl(q\to \,v\, \biggr)\ .
\label{intpol}
\eqe
If $|q| \ge \ka\,$ we find 
\[
\int_{{\ka \over |q|}}^1
\!{d\tau \over \tau }\
\log^{\la}\bigr(\sup({|\vec{p}^{\;\tau}| \over \ka},{\ka\over
  m})\bigl)\ \le \
\int_{{\ka \over |\vec{p}|}}^1
\!{d\tau \over \tau }\
\log^{\la}\bigr ({\tau |\vec{p}|\over \ka}\bigl)\ \le \
\int^{{ |\vec{p}|\over\ka }}_1
\!{dx \over x }\
\log^{\la}x \ = \frac{\log^{\la+1}({|\vec{p}| \over \ka})}{\la+1} 
\]
with an analogous calculation for  $|v| \ge \ka\,$. 
We thus obtain  a bound for (\ref{intpol})
\eq
2\,\log^{\la}\bigl(\sup({|\vec{p}| \over \ka},{\ka\over m})\bigr) \,+\, 
2\, 
\frac{\log^{\la+1}\bigl(\sup(1,{|\vec{p}| \over \ka})\bigr)}{\la+1}
\label{41}
\eqe
which allows to bound (\ref{2term}) by
\eq
\frac{6\,K^{2l-\frac14}}{(l+1)^2\, 2^4}\
(1+l)! \ \sum_{\la=0}^{\la =l}
\frac{1}{2^{\la}\,\la!}\ 
\log^{\la}\bigr(\sup({|\vec{p}| \over \ka},{\ka\over
  m})\bigl)\ .
\label{bd2term}
\eqe
Using this bound together with the previous one on 
$\, {\cal L}^{\La,\Lao}_{4,l}(0,0,0)\,$ we verify the induction hypothesis on
$\,{\cal L}_{4,l}^{\La,\Lao}(0,q,v)\,$ (\ref{prop30}) 
under the condition
\eq
K^{-1} \Bigl( 6\   K_2\ K_3\ ({6 \atop 2})\ \frac{2^4}{2\cdot 3^4}\
\frac{\,(l+1)^2 }{l^2} \ +\
16\ K''_0\ K_1 \Bigr)\ +\
6\ K^{-1/4} \le \ 1\ .
\label{bdk10}
\eqe
\\[.1cm] 
\noindent
ii)  We assume without loss of generality
$ \inf_i \eta_{i,4}^{(4)}\,=\,|p_1|\,$.
We use again an integrated Taylor formula along the integration path 
$(p_1^{\tau},\,p_2^{\tau},\,p_3^{\tau}\,)=\,
(\tau\,p_1,\,p_2,\,p_3+(1-\tau)\,p_1\,)\,$.
By Lemma 8 we find $\,
\eta_{1,4}^{(4)}(\tau) =|p_1^{\tau}|=\, \tau |p_1|$, $ 
\eta_{3,4}^{(4)}(\tau)\ge \tau |p_1|\,$.
The boundary term for $\tau=\,0\,$ is bounded  in i).
For the second term we bound
\[
|\sum_{\mu}\int_0^1d\tau\,\Bigl(
p_{1,\mu}\,\bigl(\pa_{p_{1,\mu}}-\,\pa_{p_{3,\mu}}\bigr)
{\cal L}(p_1^{\tau},\,p_2^{\tau},\,p_3^{\tau}\,)\Bigl)|
\]
\[
\le\ 
\frac{K^{2l-\frac14}(1+l)!}{(l+1)^2\, 2^4}
 \sum_{\la=0}^{\la =l-1}
\frac{|p_{1}| }{2^{\la}\,\la!}\ 
\int_0^1d\tau\,
\bigl( { 1\over  \sup(\ka, \eta_{1,4}^{(4)}(\tau)\,)}\,+\,
{ 1\over  \sup(\ka, \eta_{3,4}^{(4)}(\tau)\,)}\bigr)
\log^{\la}\bigr(\sup({|\vec{p}^{\;\tau}| \over \ka},{\ka\over
  m})\bigl)
\]
\[
\le
\frac{K^{2l-\frac14}(1+l)!}{(l+1)^2\, 2^4}
 \sum_{\la=0}^{\la =l-1}
\frac{2\,|p_{1}| }{2^{\la}\,\la!}\ 
\,\biggl(\int_0^{\inf(1,{\ka \over |p_1|})} 
{ d\tau \over \ka}\,+\,
 \int_{\inf(1,{\ka \over |p_1|})}^1  
{ d\tau\over \tau |p_1|}\biggr)\,
\log^{\la}\bigr(\sup({|\vec{p}^{\;\tau}| \over \ka},{\ka\over
  m})\bigl)
\]
\[
\le \
2\ \frac{K^{2l-\frac14}(1+l)!}{(l+1)^2\, 2^4}
 \sum_{\la=0}^{\la =l-1}
\frac{1 }{2^{\la}\,\la!}\ 
\bigr[1\,+\,\log\bigr(\sup(1,{|p_1|\over \ka})\bigl)\bigl]\
\log^{\la}\bigr(\sup({|\vec{p}^{\;\tau}| \over \ka},{\ka\over
  m})\bigl)
\]
\eq
\le\ \frac{2\,K^{2l-\frac14}}{(l+1)^2\, 2^4}\
(1+l)! \ \sum_{\la=0}^{\la =l-1}
\frac{1}{2^{\la}\,\la!}\ 
\log^{\la}\bigr(\sup({|\vec{p}| \over \ka},{\ka\over
  m})\bigl)\ 
\Bigl(1+\log \bigr(\sup(\frac{\ka}{m},{|\vec{p}| \over \ka})\bigr)\Bigr)\ .
\label{bdiiterm}
\eqe
Adding the terms from i) 
to this term gives the lower bound on $K$ 
\eq
K^{-1} \Bigl( 2\    K_2\ K_3\ ({6 \atop 2})\ \frac{2^4}{2\cdot 3^4}\
\frac{\,(l+1)^2 }{l^2} \ +\
  16\ K''_0\ K_1 \Bigr)\ +\
4\ K^{-1/4} \le \ 1\  .
\label{bdk11}
\eqe
Here we used the fact that we may bound the term from i) also by
(\ref{bdiiterm}) instead of (\ref{bd2term}) if we only want to verify the
weaker form of the induction hypothesis
valid for general momenta. At the same time we have replaced a factor
of 6 appearing in  (\ref{bdk8}) by a factor of 2, since in the general
case we may use the second bound in (\ref{57}). 
\\[.1cm]
iii)  We choose the integration path 
$(p_1^{\tau},\,p_2^{\tau},\,p_3^{\tau}\,)=\,
(\tau\,p,\,-p,\,v\,)$.
Here we assume without restriction that 
$\,|v| \le |v-(1-\tau)p|\,$, otherwise we could interchange the role of
$v\,$ and $p_4=-v\,$.
The boundary term leads again back to i).
The integral $\int_0^{1} d\tau\,$ 
is cut into four pieces - where the configuration $\ka < 2 |p_1|\,$
gives the largest contribution~: 
\[
\int_0^{1}\,=\,
\int_0^{\inf(1/2,{\ka \over |p_1|})}\,+\,
\int_{\inf(1/2,{\ka \over |p_1|})}^{1/2}\,+\,
\int_{1/2}^{\sup(1/2,1-{\ka \over |p_1|})} \,+\,
\int_{\sup(1/2,1-{\ka \over |p_1|})}^1\ .
\]
They are bounded in analogy with ii) using 
$\,\eta_{1,4}^{(4)}(\tau\,) =\, \tau |p_1|\,$ for $\,\tau \le 1/2\,$,
$\,\eta_{1,4}^{(4)}(\tau\,) =\,(1- \tau) |p_1|\,$ for 
$\,\tau \ge 1/2\,$, relations established with the aid
of Lemma 8. We get the bound
\eq
\frac{K^{2l-\frac14}}{(l+1)^2\, 2^4}\
(1+l)! \ \sum_{\la=0}^{\la =|l-1|}
\frac{1}{2^{\la}\,\la!}\ 
\log^{\la}\bigr(\sup({|\vec{p}| \over \ka},{\ka\over
  m})\bigl)\ 
\Bigl(1+2 \log \bigr(\sup(1,{|\vec{p}| \over 2 \ka})\bigr)\Bigr)
\label{bdiiiterm}
\eqe
so that verification of  (\ref{prop3}) 
requires again the lower bound (\ref{bdk11})
on $K\,$ .\\[.1cm]
iv) We assume without loss 
$\inf_i \eta_{i,4}^{(4)}\,=\,|p_1+p_2|\,$ and integrate along
$(p_1^{\tau},\,p_2^{\tau},\,p_3^{\tau}\,)=\,
(p_1,\,-p_1+\tau(p_1+p_2),\,p_3\,)$.
The boundary term has been bounded in iii).
Using Lemma 8 we find
$\,\inf \eta_{2,4}^{(4)}(\tau)=\,\tau |p_1+p_2|\,$,
and the integration term is then bounded through
\[
|\sum_{\mu}\int_0^1d\tau\,\Bigl(
(p_{1,\mu}+p_{2,\mu}) \,\pa_{p_{2,\mu}}
{\cal L}^{\La,\Lao}(p_1^{\tau},\,p_2^{\tau},\,p_3^{\tau}\,)\Bigl)|\le\,
\]
\[
\frac{K^{2l-\frac14}(1+l)!}{(l+1)^2\, 2^4}
 \sum_{\la=0}^{\la =l-1}
\frac{|p_{1}+p_{2}|}{2^{\la}\,\la!} 
\biggl(\int_0^{\inf(1,{\ka\over |p_1+p_2|})}\! \!{d\tau \over \ka}+\,
 \int_{\inf(1,{\ka\over |p_1+p_2|})}^1 \!  
{d\tau \over \tau |p_1+p_2|}\biggr)
\log^{\la}
(\sup({|\vec{p}^{\tau}| \over \ka},\,{\ka \over m}))
\]
which gives  as before a bound
\eq
\frac{K^{2l-\frac14}}{(l+1)^2\, 2^4}\
(1+l)! \ \sum_{\la=0}^{\la =l-1}
\frac{1}{2^{\la}\,\la!}\ 
\log^{\la}\bigr(\sup({|\vec{p}| \over \ka},{\ka\over
  m})\bigl)\ 
\Bigl(1+\log \bigr(\sup(1,{|\vec{p}|\over \ka})\bigr)\Bigr)
\label{bdivterm}
\eqe
so that taking into account the boundary term from iii) 
 we finally require
\eq
K^{-1} \Bigl( 2\   K_2\  K_3\ ({6 \atop 2})\ \frac{2^4}{2\cdot 3^4}\
\frac{\,(l+1)^2 }{l^2} \ +\
   16\ K''_0\ K_1 \Bigr)\ +\
5\ K^{-1/4} \le \ 1
\label{bdk12}
\eqe
to be in agreement with induction.
\\[3.5cm]
b) $2n=2\,$~:\\[.1cm]
We again use the simplified notation  (\ref{sl2}) to (\ref{sl0}).
We will assume that $\,l\ge 2\,$. 
We proceed in descending order of $|w|\,$ starting from\\[.2cm]
b1) $|w|=2\,$~:\\
\eq
\pa^{2} {\cal L}_{2,l}(p)=\,
\pa^{2} {\cal L}_{2,l}(0)\,+\,
p \int_0^1 d\tau \
\pa^3 {\cal L}_{2,l}(\tau p) \ .
\label{sloe}
\eqe
We first look at 
$\, \pa^{2}{\cal L}_{2,l}^{\La,\Lao}(0)\,$ which is decomposed as
\[
\pa^{2} {\cal L}_{2,l}^{\La,\Lao}(0)\,=\,
\pa^{2} {\cal L}_{2,l}^{0,\Lao}(0)\,+\,
\int_0^{\La} d\La'\ \pa_{\La'} \pa^{2} {\cal L}_{2,l}^{\La',\Lao}(0)\ ,
\]
the second term being obtained from the r.h.s. of the FE,
and the first vanishing by (\ref{ren}).
The {\it first} term on the r.h.s. of the FE then gives the bound
\[
({4 \atop 2}) \
\int_m^{\La+m} d\ka' \int_k \frac{2}{\La'^3}\ e^{-{{k^2+m^2}\over \La'^2}}
\ \ka'^{-2}\ \frac{K^{2l-2-\frac14}}{l^2\, 2^4}\ 
\ l!\ \sum_{\la=0}^{\la =l-2}  
\frac{\log^{\la}(\sup({|k| \over \ka'},{\ka'\over m}))}{2^{\la}\,\la!}
\]
\eq
\leq \  K_2\ K_3\ \frac{6}{2^4}\ \frac{K^{2l-2-\frac14}}{l^2} \
\ l!\ \sum_{\la=0}^{\la =l-2}
\frac{1}{2^{\la}\,\la!}\
\int_m^{\La+m} d\ka'\ \ka'^{-1} \ 
\Bigl(\,\log^{\la}({\ka'\over m})
\,+\,   (\la!)^{1/2}\,\Bigr)\ ,
\label{1stinr2}
\eqe
where we used again (\ref{momint}) and (\ref{la3}), remembering that
$\,|\vec{p}|=k\,$ in the present case. 
Using (\ref{wehre}) and (\ref{57}) (with $l \to l-1\,$) 
the first term on the
r.h.s. of the FE is then bounded
in agreement with the induction hypothesis by
\[
\frac{K^{2l-1}}{(l+1)^2\, 2^4}\
l!\ \sum_{\la=0}^{\la =l-1}
\frac{\log^{\la}({\ka\over m})}{2^{\la}\,\la!}
\] 
under the assumption
\eq
K^{-5/4} \  K_2\ K_3 \ 6\cdot 6\
\frac{\,(l+1)^2 }{l^2} \ \le  \ 1\   . 
\label{bdk13}
\eqe
This contribution has to be added to the one from the {\it second} term
on the r.h.s. of the FE.
We have only contributions with 
$n_1=1$ and $n_2=1$.  The two momentum derivatives have to apply  both
to the propagator  or both
to a function ${\cal L}_{2,l}$~; all
other contributions vanish at zero momentum.
For the contribution of the first kind 
integration of the FE (\ref{feq}) gives the   bound
\[
8\int_0^{\La}\frac{ d\La'}{\La'^{5}}\  e ^{-\frac{m^2}{\La '^2}}
\ \ka'^{4}\ K^{2l-2}\! \! \!  \! \! \! 
\sum_{\begin{array}{c}_{l_1+l_2=l},\\[-.2cm]_{l_1,l_2 \ge 1}\end{array} }  
\frac{ l_1!}{(l_1+1)^2} \ \frac{l_2 !}{(l_2+1)^2\, } 
\sum_{\la_1=0}^{\la_1 =l_1-1}
\frac{\log^{\la_1}({\ka '\over m})}{2^{\la_1}\,\la_1!}\
\sum_{\la_2=0}^{\la =l_2-1} 
\frac{\log^{\la_2}({\ka'\over m}))}
{2^{\la_2}\,\la_2!}
\]
\eq
\le\
8\,\frac{K^{2l-2}\,  K''_0}{(l+1)^2}\ 
\int_0^{\La} \frac{ d\La'}{\La'^{5}}\ 
e ^{-\frac{m^2}{\La '^2}}\ \ka'^{4} 
\sum_{\la=0}^{\la =l-2}
\frac{\log^{\la}({\ka '\over m})}{2^{\la}\,\la!}\ ,
\label{wo}
\eqe
where we used (\ref{34})and 
Lemma 2 c). Using also Lemma 7
we obtain the bound
\eq
16\ \frac{K^{2l-2}\,  K_0''}{(l+1)^2}\ 
K_1'\ 
\sum_{\la=1}^{\la =l-1}
\frac{\log^{\la}({\ka \over m})}{2^{\la}\,\la!}\ .
\label{1w2}
\eqe
For the contribution of the second kind 
integration of the FE gives in the same way the bound 
(again using Lemma 2 c) and Lemma 7)
\[
4\,\int_0^{\La} \frac{d\La'}{\La'^{3}}\ e ^{-\frac{m^2}{\La '^2}}\ \ka'^{2}\
K^{ 2l-2}\!\! \!   
\sum_{\begin{array}{c}_{l_1+l_2=l},\\_{l_1,l_2 \ge 1}\end{array} }  
\frac{ l_1!}{(l_1+1)^2} \ \frac{l_2 !}{(l_2+1)^2\, } 
\sum_{\la_1=0}^{\la_1 =l_1-1}
\frac{\log^{\la_1}({\ka '\over m})}{2^{\la_1}\,\la_1!}\
\sum_{\la_2=0}^{\la =l_2-1} 
\frac{\log^{\la_2}({\ka'\over m})}
{2^{\la_2}\,\la_2!}
\]
\eq
\le\
4\ \frac{K^{ 2l-2}\,  K_0''}{(l+1)^2}\
\int_0^{\La} \frac{d\La'}{\La'^{3}}\ 
e ^{-\frac{m^2}{\La '^2}}\ \ka'^{2} 
\sum_{\la=0}^{\la =l-2}
\frac{\log^{\la}({\ka '\over m})}{2^{\la}\,\la!}\
\le \ 8\ \frac{K^{ 2l-2}\,  K_0''}{(l+1)^2}\ 
K_1\ 
\sum_{\la=1}^{\la =l-1}
\frac{\log^{\la}({\ka \over m})}{2^{\la}\,\la!}\ .
\label{woe}
\eqe
The sum of   this bound and the bounds (\ref{bdk13}), (\ref{1w2})
 is compatible with the induction hypothesis (\ref{prop6})
under the condition
\eq
K^{-5/4} \  K_2\  K_3\ 36\
\frac{\,(l+1)^2 }{l^2} \ +\
8\ K^{-1}(   2\, K_0''\ K'_1 
\ +\ K_0''\ K_1)   \ \le \ 1\ . 
\label{bdk15}
\eqe 
The second term in (\ref{sloe})  is bounded
with the aid of the induction hypothesis
\[
|p| \int_0^1 d\tau \, 
\pa^{3} {\cal L}_{2,l}(\tau p)|
\ \leq\
|p| \int_0^1 { d\tau\over 
\sup(\tau|p|,\ka)}\
\frac{K^{2l-1-\frac14}}{(l+1)^2}
 \ l!\ \sum_{\la=0}^{\la=l-2}\frac{1}{2 ^{\la}\,\la!}\
\log^{\la}\bigr(\sup({|\tau\,p| \over \ka},{\ka\over m})\bigl)\ .
\]
Assuming that $|p| > \ka\,$ 
and also that $|p|\,m > \ka ^2 \,$, which is the most delicate
case (in the other cases some of the 3 contributions in (\ref{astt})
below are absent) we cut up the integral 
\[
\ \int_0^1  d\tau \ =\ \Bigl( \int_0^{\frac{\ka}{p}}\,+\,
\int_{\frac{\ka}{p}}^{\frac{\ka ^2}{pm}}\,+\, 
\int_{\frac{\ka ^2}{pm}}^1\Bigr) 
d\tau
\]
and find
\eq
|p| \Bigl(\int_0^{\frac{\ka}{p}}+
\int_{\frac{\ka}{p}}^{\frac{\ka ^2}{pm}}+ 
\int_{\frac{\ka ^2}{pm}}^1\Bigr)  d\tau
{\log^{\la}\bigr(\sup({|\tau\,p| \over \ka},{\ka\over m})\bigl)
\over 
\sup(\tau|p|,\ka)}\,\le\,
\log^{\la}({\ka\over m})
\,+\,\log^{\la+1}({\ka\over m})
\,+\,\frac{\log^{\la+1}({p\over \ka})}{\la +1}
\label{astt}
\eqe
so that we obtain the bound
\[
2\  \frac{K^{2l-1-\frac14}}{(l+1)^2}\
l!\  \sum_{\la=0}^{\la=l-2}\frac{1}{2 ^{\la}\,\la!}\
\log^{\la}({\ka\over m}))
\Bigl(1+\log(\sup({|p| \over \ka},\frac{\ka}{m}))\Bigr)\ .
\]
The final lower bound on $K\,$ is obtained on adding the bound  
(\ref{bdk15}) stemming from the boundary term  at zero momentum 
and this one 
\eq
2\, K^{-1/4}\,+\,K^{-5/4} \  K_2\  K_3\ 12\
\frac{\,(l+1)^2 }{l^2} \ +\
8\ K^{-1}(   2\, K_0''\ K'_1 
\ +\  \, K_0''\ K_1)   \ \le \ 1\ .
\label{bdk16}
\eqe 
In the second term we again replaced a factor of 6 by a factor of 2
as in (\ref{bdk11}).\\[.1cm]
\noindent
b2) $|w| = 1\,$~:\\
In this case we write 
\eq
\pa {\cal L}_{2,l}(p)=\,
\pa {\cal L}_{2,l}(0)\,+\,
p \,\pa^2 {\cal L}_{2,l}(0) \,+\,
p^2 \int_0^1 d\tau \ (1-\tau)\ \pa^3
{\cal L}_{2,l}(\tau p) \ .
\label{sloe1}
\eqe
Due to Euclidean symmetry the first term on the r.h.s.
vanishes. The bound on the second term
has been calculated in the previous section.
The last term is bounded as in the previous
calculation by  
\[
2\ \sup(p,\ka)\ 
 \frac{K^{2l-1-\frac14}}{(l+1)^2}\
l!\  \sum_{\la=0}^{\la=l-2}\frac{1}{2 ^{\la}\,\la!}\
\log^{\la}({\ka\over m}))
\Bigl(1+\log(\sup({|p| \over \ka},\frac{\ka}{m}))\Bigr)\ .
\]
so that we get again the lower bound (\ref{bdk16}) on $\,K\,$.\\[.1cm]
b3) $|w| = 0\,$~:\\
We first look at 
$\, {\cal L}_{2,l}^{\La,\Lao}(0)\,$ which is written as
\eq
 {\cal L}_{2,l}^{\La,\Lao}(0)\,=\,
 {\cal L}_{2,l}^{0,\Lao}(0)\,+\,
\int_0^{\La} d\La'\ \pa_{\La'}  {\cal L}_{2,l}^{\La',\Lao}(0)\ .
\label{2pt}
\eqe
From the  {\it first} term on the r.h.s. of the FE,
where we use the bound (\ref{prop30}) since two of the external
momenta
in $ {\cal L}_{4,l-1}^{\La',\Lao}(0,0,k,-k)\,$ vanish, we obtain
using again (\ref{la3}) and (\ref{momint})
\[
({4 \atop 2}) \
\int_m^{\ka} d\ka' \int_k \frac{2}{\La'^3}\ e^{-{{k^2+m^2}\over \La'^2}}
\ \frac{K^{2l-2}}{l^2\, 2^4}\ 
\ l!\ \sum_{\la=0}^{\la =l-1}  
\frac{\log^{\la}(\sup({|k| \over \ka'},{\ka'\over m}))}{2^{\la}\,\la!}
\]
\[
\leq \   \frac{6}{2^4}\ K_2\ K_3\ \frac{K^{2l-2}}{l^2} \
l!\ \sum_{\la=0}^{\la =l-1}
\frac{1}{2^{\la}\,\la!}\
\int_m^{\ka} d\ka'\ \ka' \ 
\Bigl(\,\log^{\la}({\ka'\over m})\,+\,  
(\la!)^{1/2}\,\Bigr)
\]
\eq
\leq \   \frac{6\, K_2\,  K_3}{2^4}\ \frac{K^{2l-2}}{l^2} \
l!\ \sum_{\la=0}^{\la =l-1}
\frac{1}{2^{\la}\,\la!}\
\frac{\ka^2}{2} \ 
\Bigl(\,\log^{\la}({\ka\over m})\,+\,  
(\la!)^{1/2}\,\Bigr)
\label{1stinr22}
\eqe
\eq
\leq \  3\  \frac{6\, K_2\,  K_3}{2^4}\ \frac{K^{2l-2}}{l^2} \
l!\ \sum_{\la=0}^{\la =l-1}
\frac{1}{2^{\la}\,\la!}\
\frac{\ka^2}{2} \ 
\log^{\la}({\ka\over m})\ .
\label{1stinr3}
\eqe
This is compatible with the induction hypothesis (\ref{prop6}) if
\eq
K \ge \frac{9}{2^4}\ \frac{(l+1)^2}{l^2}\  K_2\,  K_3\ .
\label{kal}
\eqe
Integrating the {\it second} term on the r.h.s. of the FE we obtain the bound 
\[
4\int_0^{\La} \frac{d\La'}{\La'^{3}}\  e ^{-\frac{m^2}{\La '^2}}\ \ka'^{4}\
K^{2l-2}\!\!\!\!\!  
\sum_{\begin{array}{c}_{l_1+l_2=l},\\[-.2cm]_{l_1,l_2 \ge 1}\end{array} }  
\frac{ l_1!}{(l_1+1)^2} \ \frac{l_2 !}{(l_2+1)^2\, } 
\sum_{\la_1=0}^{l_1-1}
\frac{\log^{\la_1}({\ka '\over m})}{2^{\la_1}\,\la_1!}\
\sum_{\la_2=0}^{l_2-1} 
\frac{\log^{\la_2}({\ka'\over m}))}
{2^{\la_2}\,\la_2!}
\]
\[
\le\
4\ \frac{K^{2l-2}\,  K_0''}{(l+1)^2}\ 
\int_0^{\La} \frac{d\La'}{\La'^{3}}\ 
e ^{-\frac{m^2}{\La '^2}}\ \ka'^{2} 
\sum_{\la=0}^{\la =l-2}
\frac{\log^{\la}({\ka '\over m})}{2^{\la}\,\la!}
\ \le\ 
\ka ^2\ \frac{K^{2l-1}}{(l+1)^2}\ l!
\sum_{\la=1}^{\la =l-1}
\frac{\log^{\la}({\ka \over m})}{2^{\la}\,\la!}
\]
using again Lemma 2c) and Lemma 7 and imposing the condition
\eq
4 \ K_0'' K_1 \ \le\ K \ .
 \label{bdk17}
\eqe
\noindent
To go away from zero momentum we write similarly as in (\ref{sloe1})
\eq
{\cal L}_{2,l}(p)=\,
{\cal L}_{2,l}(0)\,+\,
\frac12\ p^2 \,\pa^2 {\cal L}_{2,l}(0) \,+\,
p^3 \int_0^1 d\tau\ \frac{(1-\tau)^2}{2!}\ \pa^3
{\cal L}_{2,l}(\tau p) 
\label{sloe2}
\eqe
and proceed in the same way as  in the previous section, see
(\ref{bdk13}), (\ref{wo}), (\ref{woe}),   (\ref{astt}).
Inductive verification of (\ref{prop5}) gives 
similarly as in (\ref{bdk16}) the lower bound on $\,K\,$ 
\eq
K^{-\frac14}+K^{-\frac54}   K_2\ K_3\ 6\
\frac{(l+1)^2 }{l^2}  +
K^{-1}\Bigl(\frac12\ \frac{9}{2^4}\ \frac{(l+1)^2}{l^2}\   K_2\,  K_3 +
6 \ K_0'' K_1  +
8 \ K_0'' K'_1 \Bigr)    \le 1
\label{bdk18}
\eqe 
noting that factors of $1/2\,$ are gained since
\[
\sum_{\la=1}^{\la =l-1}
\frac{\log^{\la}({\ka \over m})}{2^{\la}\,\la!}
\ \le \
\frac12\ \sum_{\la=0}^{\la =l-2}
\frac{\log^{\la}({\ka \over m})}{2^{\la}\,\la!}\
(1\,+\,\log^{\la}({\ka \over m}))\ .
\]
\qed

\noindent
 \section*{References}

\begin{itemize}
\item[{[Br]}] S. Breen, Leading large order
  asymptotics for $(\phi^4)_2$ perturbation Theory, 
Commun. Math. Phys. {\bf 92}, 179-192 (1983).  
\item[{[CPR]}] C. de Calan, D. Petritis and V. Rivasseau,
Local Existence of the Borel Transform in Euclidean $\phi_4^4$,
Local Existence of the Borel Transform in Euclidean Massless $\phi_4^4$,
Commun. Math. Phys. {\bf 101}, 559-577 (1985).  
\item[{[CR]}] C. de Calan and V. Rivasseau,
Local Existence of the Borel Transform in Euclidean $\phi_4^4$,
Commun. Math. Phys. {\bf 82}, 69-100 (1981).  
\item[{[DFR]}]  F. David, J. Feldman  and V. Rivasseau,
On the Large Order Behaviour of $\phi_4^4$,
Commun. Math. Phys. {\bf 116}, 215-233 (1988).
\item[{[FHRW]}]  J. Feldman, T. Hurd, L. Rosen, J. Wright,
QED~: A proof of Renormalizability, Lecture Notes in Physics,
Vol. {\bf 312}, Springer-Verlag 1988.
\item[{[FMRS]}]  J. Feldman, J. Magnen, V. Rivasseau and R. S\'en\'eor,
Bounds on Renormalized Feynman Graphs,
Commun. Math. Phys. {\bf 100}, 23-55 (1985).
\item[{[GK]}] R. Guida, Ch. Kopper, Large momentum bounds for massless
$\Phi^4_4$, to appear.
\item[{[Ke]}] G. Keller, Local Borel summability of Euclidean
$\Phi^4_4$: A simple Proof via Differential Flow Equations.
Commun. Math. Phys. {\bf 161}, 311-323 (1994).
\item[{[KKS]}] G. Keller, Ch. Kopper and M. Salmhofer,
Perturbative renormalization and effective Lagrangians in $\Phi_4^4$,
Helv. Phys. Acta {\bf 156} , 32-52 (1992). 
\item[{[KM]}] Ch. Kopper and F. Meunier, 
Large momentum bounds from Flow equations 
 Ann. Henri Poincar{\'e} {\bf 3}, 435-450 (2002). 
\item[{[KMR]}] Ch. Kopper, V.F. M\"uller and Th. Reisz,
Temperature Independent Renormalization of Finite Temperature 
Field Theory, Ann. Henri Poincar{\'e} {\bf 2}, 
387-402 (2001).
\item[{[KM\"u]}]  Ch. Kopper and V.F. M\"uller,
Renormalization  Proof for  Massive $\varphi_4^4$-Theory 
on Riemannian Manifolds, Commun.Math.Phys. {\bf 275}, 331-372 (2007). 
\item[{[Ko]}]  Ch. Kopper, 
Continuity of the four-point function
of massive $\varphi_4^4$-Theory above threshold,
Rev. Math. Phys. {\bf 19}, 725-747 (2007).
\item[{[Li]}] L. N. Lipatov, Divergence of the perturbation theory
series and quasi-classical theory, Sov. Phys. JETP {\bf 45}, 216-223 (1977). 
\item[{[MNRS]}] J. Magnen, F. Nicol\`o, V. Rivasseau, R. S\'en\'eor,
Commun. Math. Phys. {\bf 108}, 257-289 (1987).  
\item[{[MR]}] J. Magnen, V. Rivasseau, The Lipatov Argument for 
$\phi_3^4$ Perturbation Theory,
Commun. Math. Phys. {\bf 102}, 59-88 (1985).  
\item[{[M\"u]}] V.F. M\"uller, Perturbative Renormalization by Flow
  Equations,\\ Rev. Math. Phys. {\bf 15}, 491-557 (2003).
\item[{[Po]}] J. Polchinski: Renormalization and Effective Lagrangians,
Nucl.Phys.{\bf B231}, 269-295 (1984).
\item[{[Ri]}] V. Rivasseau, Construction and Borel Summability of
  Planar 4-Dimensional Euclidean Field Theory,
Commun. Math. Phys. {\bf 95}, 445-486 (1984).  
\item[{[Sp]}] T. Spencer, The Lipatov argument,
Commun. Math. Phys. {\bf 74}, 273-280 (1980).  
\item[{[tH]}] G. 't Hooft, Can we make sense out of 
"Quantum Chromodynamics", in~: The Whys of Subnuclear Physics,
Proceedings of the Erice Conference 1977~; A. Zichichi ed.,
Plenum Press, New York 1979.
\item[{[WH]}] F. Wegner, A. Houghton: Renormalization Group Equations
  for Critical Pheno\-mena, Phys. Rev. {\bf  A8}, 401-412 (1973). 
\item[{[Wi]}]  K. Wilson: Renormalization group and critical phenomena 
I. Renormalization group and the Kadanoff scaling picture,
Phys.Rev. {\bf B4}, 3174-3183 (1971),\\ 
K. Wilson: Renormalization group and critical pheno\-mena
II. Phase cell analysis of
critical behaviour, Phys.Rev. {\bf B4}, 3184-3205  (1971).

\end{itemize}

\end{document}